\documentclass[aps,prd,onecolumn,12pt,preprintnumbers]{revtex4-1}

\usepackage{hyperref}

\usepackage{float}

\usepackage{tikz}

\usepackage{graphicx}
\usepackage{epstopdf}

\begin{document}

\preprint{HDP: 17 -- 02}

\title{Banjo Drum Physics --- theoretical preliminaries}

\author{David Politzer}

\email[]{politzer@theory.caltech.edu}

\homepage[]{http://www.its.caltech.edu/~politzer}

%\thanks{452-48 Caltech, Pasadena CA 91125}
\altaffiliation{\footnotesize Pasadena CA 91125}
%\altaffiliation{\newline \em \em \em 452-48 Caltech, Pasadena CA 91125}
\affiliation{}

%\date{\today}
\date{August 31, 2017 --- Helmhotz's $196^{\text{th}}$ birthday}

\begin{figure}[h!]
\includegraphics[width=4.3in]{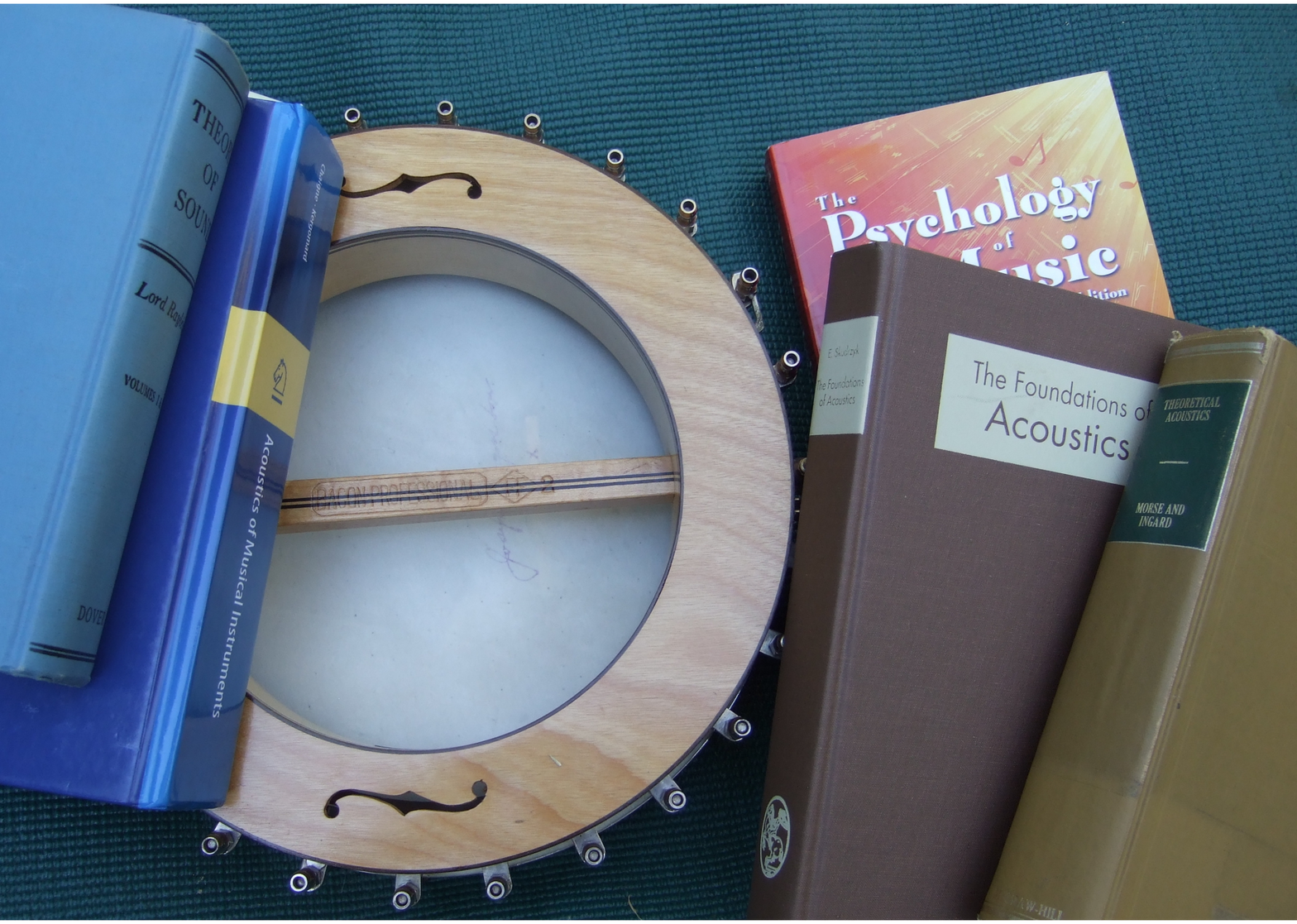}
\end{figure}

\begin{abstract}
The interaction of a drum's head with its enclosed air is presented in the simplest possible form appropriate to the questions and issues that arise in understanding the timbre of the banjo.  The inherent air-head impedance mismatch allows treating the head as driver of the air and the air's effect, in turn, as back reaction.  Any particular question can then be addressed with a calculation in simple wave mechanics.  The analysis confirms and quantifies the notion that internal air resonances enhance the response of the head at its and their frequencies.  However, the details of just how are fairly complicated.
\begin{figure}[h!]
\includegraphics[width=1.85in]{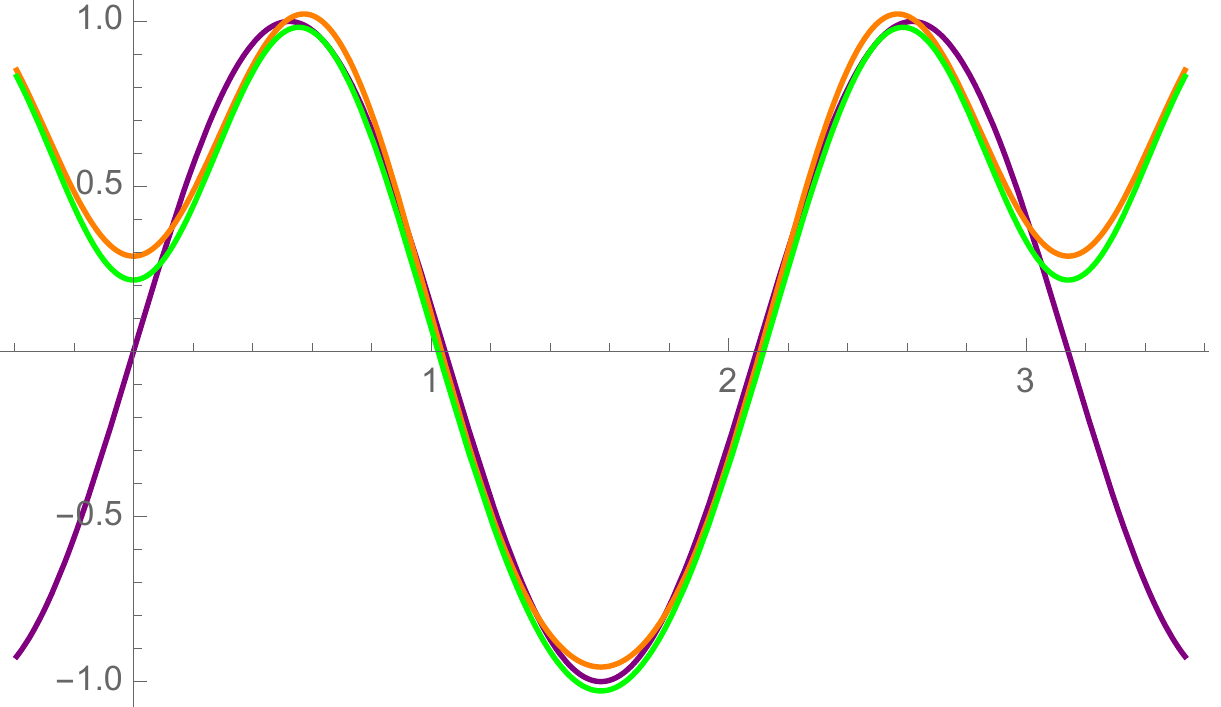}
\end{figure}
\end{abstract}

\maketitle{ {\centerline{\large \bf Banjo Drum Physics --- theoretical preliminaries}

\section{background \& goals}

Rayleigh\cite{rayleigh} may have been the first to recognize that the kettledrum (timpani) presents an interesting but tractable physics problem.  Why does it have a clearly discernible  pitch--- rather than being ``indefinite," like most drums?  The answer lies with the interaction of the enclosed air with the head.  Over subsequent decades, this problem was addressed with increasing sophistication.  The strategies have become more sophisticated, but the basic math used is still from the $19^{\text{th}}$ Century (and physics from the $18^{\text{th}}$ and $17^{\text{th}}$).  Quantitative advances have been possible because of improvements in numerical computation.

For the kettledrum, the focus is the shifts in the lowest few resonant frequencies due to the head-air interaction and the relative amplitudes of those resonances as a function of where the drummer's mallet strikes the head.  The kettledrum's lowest and strongest modes end up with nearly integer ratios --- in contrast to the distinctly non-musical ratios of the ideal drum head. 

Correct and complete solutions of the membrane-air interaction have long been known.\cite{rossing}\cite{kergomard}  The goal here is to find the most elementary approach that can give a perspective on the questions that arise for the banjo and give qualitative and potentially quantitative guides to their solutions.  Those questions are quite different from the focus of the kettledrum studies.

On a banjo, strings drive the head via the bridge and present a variety of possible frequency spectra, all quite rich in high harmonics.  Also, there are many variables in the spatial aspects of how the head is driven.  On a five-string banjo, the first and fifth strings produce the most bridge end-to-end rocking, while the third string gives no bridge rock at all.  The third string and whole bridge, however, are typically off center in the radial direction.  Both the off-center location of the bridge and its rocking contribute to excitations of the head that are not rotationally symmetric about its center.  And the footprint of the bridge on the head is a design variable.  Also important is its placement --- fixed once and for all by the builder if the neck has frets --- but potentially different for different designs.  And different pot (body) designs produce very subtle differences in timbre.   To most people, a banjo is a banjo.  However, aficionados recognize and care about the subtle distinctions.  Introductions to musical acoustics often define timbre of a sound as the collective aspect of that sound that cannot easily be quantified, e.g.~{\it not} pitch, frequency, loudness, duration, tempo, \&c.  This investigation aims at understanding subtle aspects of timbre arising from banjo pot geometries.  A variety of geometries and the whole range of audible frequencies are relevant.  And sometimes a qualitative grasp of all those possibilities might be more helpful than running computer code focused on some particulars.

\section{qualitative summary \& outline}

Previous banjo physics studies\cite{politzer} addressed the air modes themselves.  Simple theory gave a satisfactory and illuminating account of measurements for which the head was decoupled --- by replacing it with $3/4''$ plywood.  The implication was that air resonances enhanced the head motion at their frequencies and thereby altered the timbre.  The present note describes a way to make that connection more precise and to understand the head-air interaction in terms of basic physics.  Rather than precision determinations of a few parameters, what is wanted is a general, approximate approach to any particular question that provides a well-founded qualitative picture and where the calculations could be readily performed or estimated.  

Damping is obviously essential to self-consistency, especially in a frequency domain, steady-state approach.  Vibrational amplitudes must remain small to preserve approximate linearity for the sake of the simple math and for the sake of musicality to the ear.  Furthermore, the effects of the air resonances on the head must remain relatively small, even at their maximum.  In addition, an important lesson of this endeavor is an appreciation of the very significant impact on timbre that is produced by the details of the dissipation of the head and pot air vibrational energy (to sound, heat, and other parts' motions).  With many competing effects operative at once, which ones stand out and which are washed out depends on the magnitudes of the various sorts of energy loss and radiation.  Since that dissipation is particularly hard to quantify and to model, the results of the ``theory" presented here remain rather qualitative.  

Also, setting up the questions as physics problems makes it clear that our collective ignorance of how to quantify timbre means that we don't really have specific, quantifiable questions to ask.  ``What is the effect of increasing the pot depth from $2{1\over2}''$ to $3''$?" sounds like a simple question.  But the physics answer is an account of the interactions of the motions of the various parts and their effect on the  radiative motion of the air ({\it aka} sound).  Our understanding of how this relates to musical judgments is rather primitive.  We don't really know how to translate ``ring," ``clang," ``ping," or  ``sweet," ``tubby," ``muddy,"  ``clear," \&c. into quantifiable sound characteristics.\cite{deutsch}

In the spirit of dividing ``understanding the banjo" into manageable parts, the issue of external ``air-loading" of the head is not addressed here.  The approach described for the interior air would work just as well for the outside air.  In the proposed approximation scheme, the two effects are independent and would simply add.  Both can be considered as small back reactions on the head motion because of the head-air impedance mismatch.  And exterior air-loading can be considered to be a constant feature, along with string and bridge dynamics, when we focus specifically on the enclosed air and its geometry.

 \begin{figure}[h!]
 \includegraphics[width=2.8in]{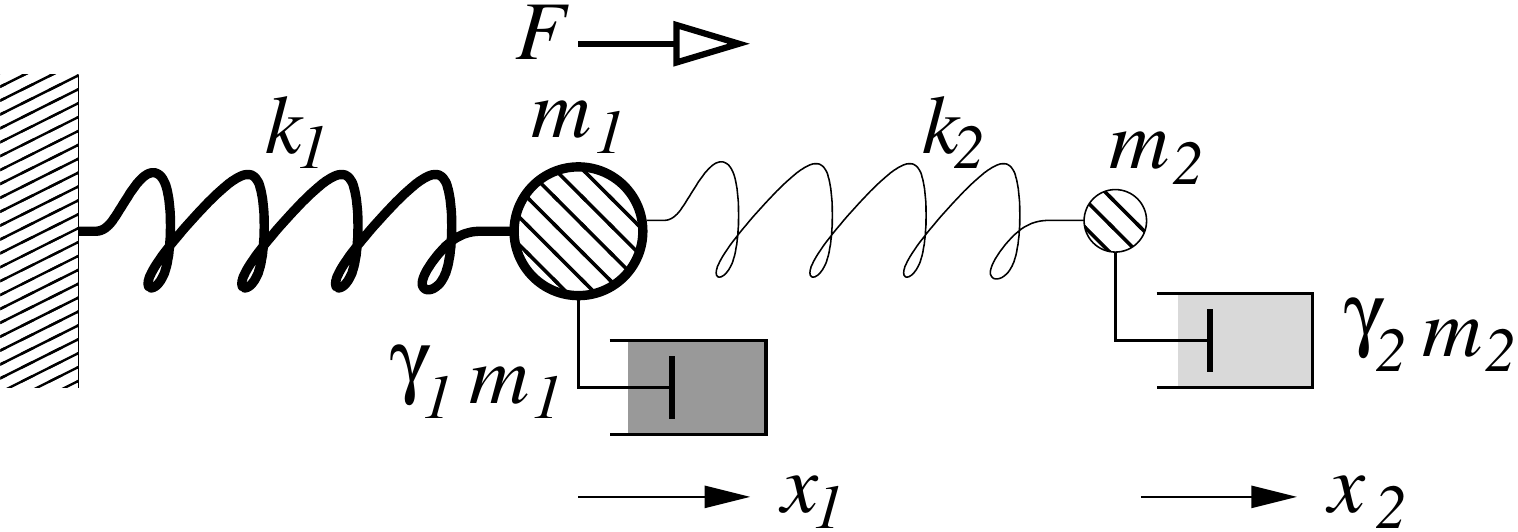}
 \caption{0-0D: two forced, coupled, damped oscillators with $m_1\gg m_2$ and $k_1\gg k_2$}
 \end{figure}

\subsection{the model systems}

I present three idealized, model systems, building up the level of complexity.  The first is just two coupled, damped oscillators.  The first oscillator is forced, and the question is the amplitude of its steady-state motion.  The oscillators by themselves have comparable frequencies, but the forced one has a much bigger mass and stiffer spring.  It serves as the model of the head, while the wimpier one is the air.  I call this system 0-0D because the position of each oscillator at any particular time is just a point, and a point is a ``space" of zero dimension.  
 \begin{figure}[h!]
 \includegraphics[width=2.6in]{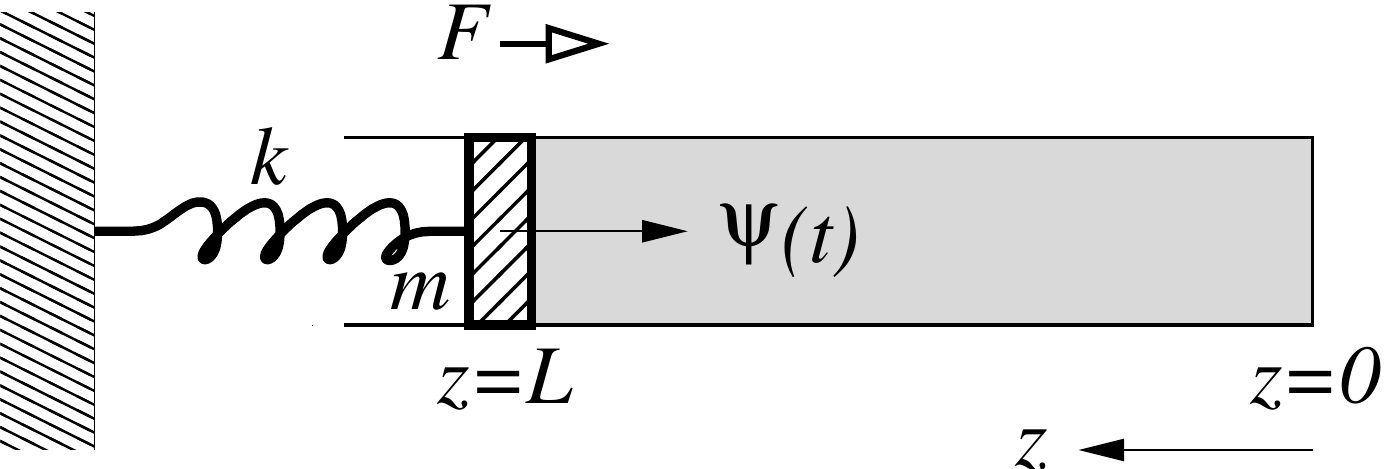}
 \caption{0-1D: forced piston on a spring with back reaction from air in cylinder}
 \end{figure}

The second system is a spring-loaded piston that moves in an air-filled cylinder.  The piston is forced, and we study its response.  This system is called 0-1D because the piston position is characterized by a single point while the air in the cylinder is characterized by the pressure and air velocity along its center line.

 \begin{figure}[h!]
 \includegraphics[width=3.4in]{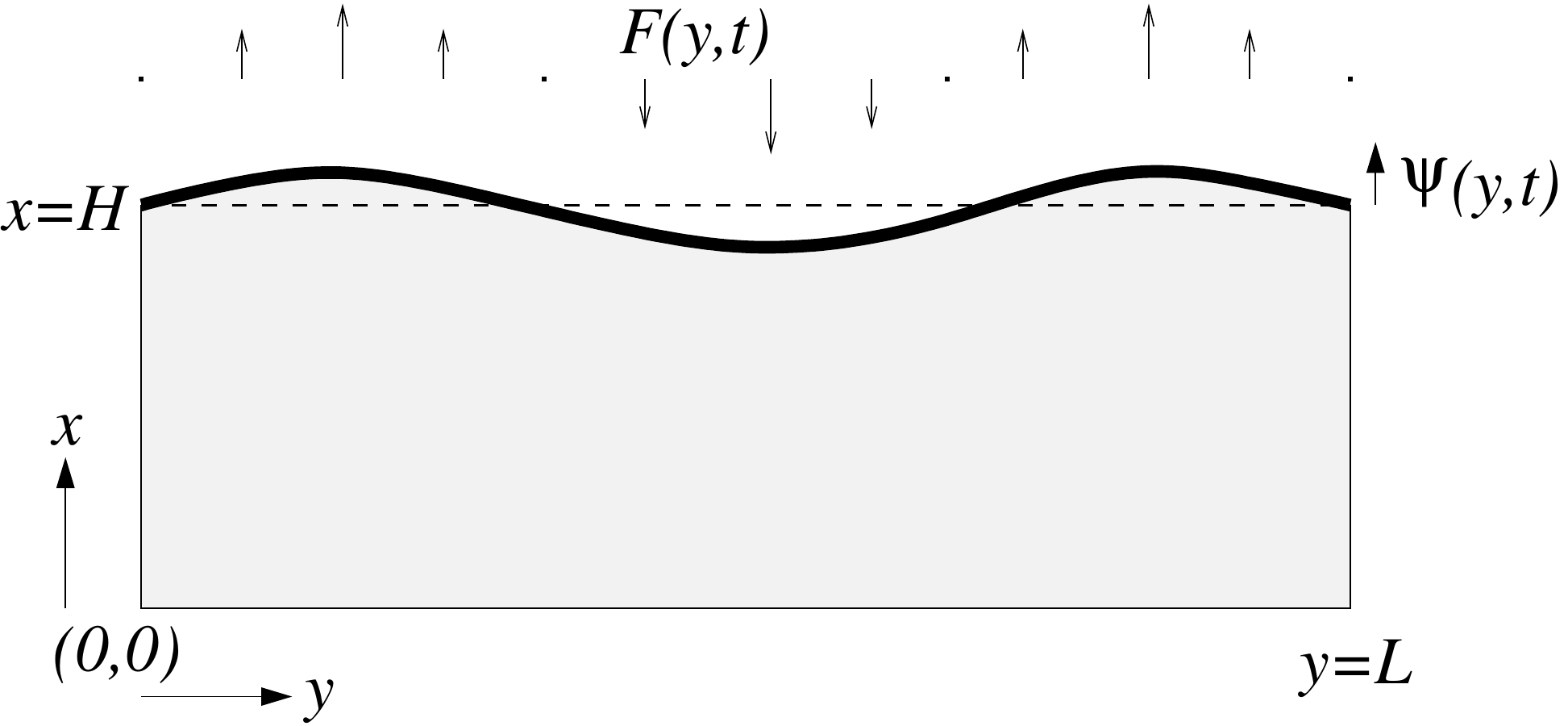}
 \caption{1-2D: specified 1D motion $\psi (y,t)$ drives 2D fluid}
 \end{figure}
  \begin{figure}[h!]
 \includegraphics[width=2.5in]{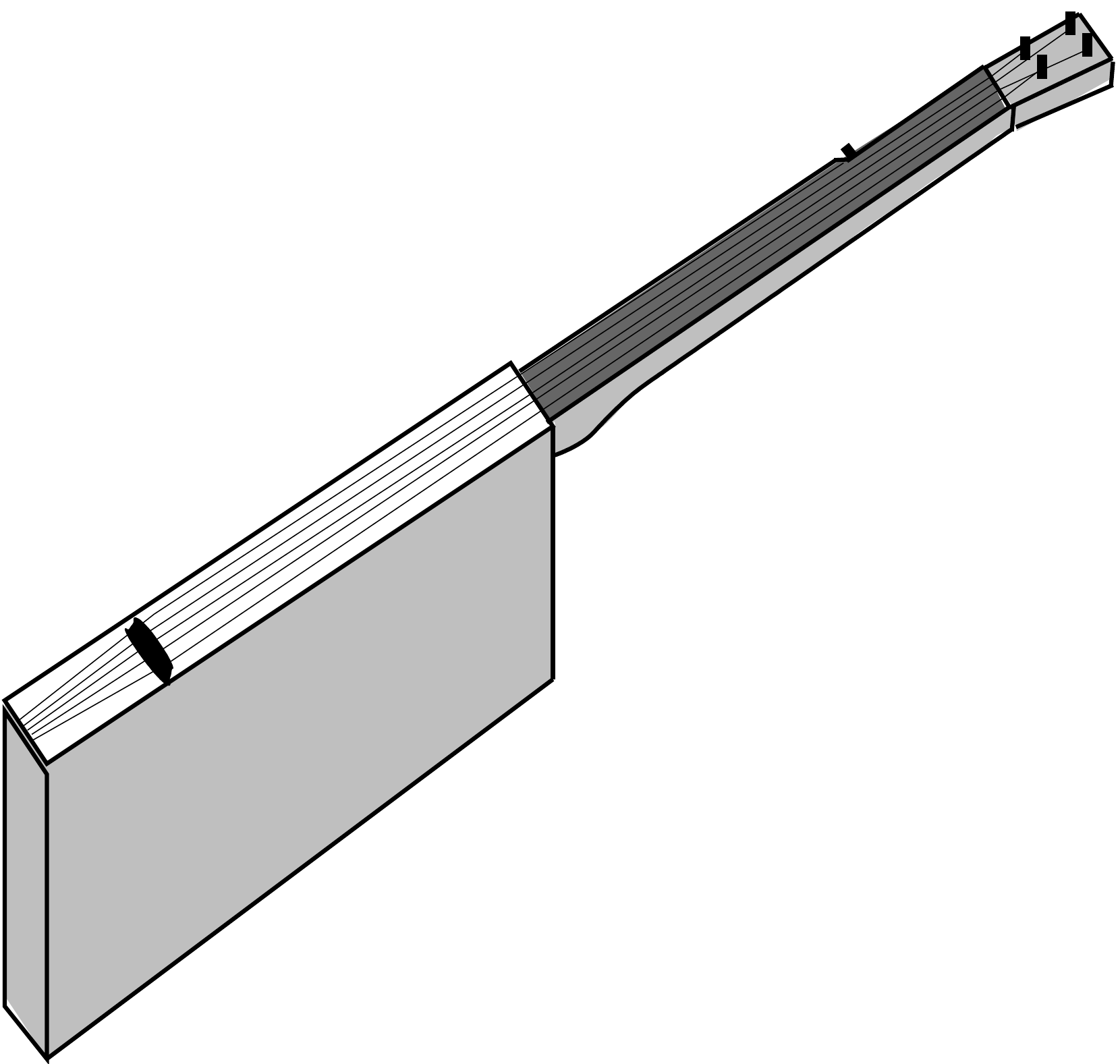}
 \caption{a physical 5-string banjo whose lower frequency behavior approximates the 1-2D system}
 \end{figure}
 
1-2D is a one-dimensional ``membrane" coupled to a two-dimensional compressible fluid.  Even restricting to a single sinusoidal frequency, there are an infinite number of ways to force the 1D membrane.  The spatial structure of the forcing is naturally decomposed into the shapes of the modes of the membrane by itself.  I look specifically at the back reaction of the fluid on the membrane, given some initial membrane mode motion.  The charming and challenging aspects of vibro- or structural acoustics make their appearance.  

An actual banjo, however idealized, would be 2-3D, and the further complications are computationally formidable.  But the essential conceptual issues are already faced in the 1-2D system.  The internal resonator provides a simple example of aspects of a particular 2-3D system.   

Typical banjos' lowest air resonance and the lowest resonance of virtually all string instruments is the Helmholtz resonance.  This requires a small port on the air cavity to the outside.  According to the standard simplified account, the enclosed air acts as a spring, and the air that oscillates in and out of the port is the mass.  On resonance, all the enclosed air expands uniformly and applies a uniform pressure over the surface of the head.  It can be excited by any head motion that can create a net volume change to the air cavity.  And the pressure reacts back on all of the head motions that involve a net volume change.  The head-air interaction can be understood as a set of 0-0D models.

\subsection{qualitative results}

The following are summaries of the qualitative results.  The equations and their solutions are given in section III.

\subsubsection{0-0D}

The two oscillator 0-0D system allows an exact solution to be compared to the leading impedance-mismatch, back reaction calculation.  The back reaction calculation proceeds as follows.  First, we solve for the steady-state motion of the forced oscillator in the absence of coupling to the second  oscillator.  Then that motion is used as a force on the second oscillator.  The consequent motion of the second oscillator is then used as the source of an additional force on the first, which gives rise to an additional motion added to the original one.  For large impedance mismatch (i.e., large difference in masses and spring constants), this approach agrees in all qualitative aspects with the exact solution.  The back reaction approach actually gives the first term in a power series expansion of the exact solution in a small parameter, e.g.~the ratio of the masses.  Subsequent terms in that expansion are simply higher order iterations.  Keeping only the first term and the expansion itself only make sense if the damping is sufficient to ensure that the back reaction force never gets large compared to the initial driving.

The uncoupled resonant frequencies are shifted slightly.  The small mass resonance appears as a bump in the driven motion of the larger one.  In the presence of the coupling, the response to a given force is enhanced at frequencies below both resonances and at frequencies above both.  The response is weakened for drive frequencies between the two resonances.  If there were no damping, there would be a symmetry of the coupling behavior immediately above and below resonance --- with an abrupt change of sign upon crossing the resonance.  Damping not only smooths the response into something continuous, it also makes the effect positive in the vicinity of the resonance, i.e.~on both sides.

FIG.s 5 and 6 assume some particular parameter values and compare the uncoupled, driven oscillator (orange or lightest), the exact solution of the two coupled oscillators (red), and the simpler calculation of driving the second oscillator with the motion of the uncoupled first and then computing and adding in the back reaction of the second onto the first (black or darkest).  See section IIIA: 0-0D for details.
\begin{figure}[h!]
 \includegraphics[width=4.5in]{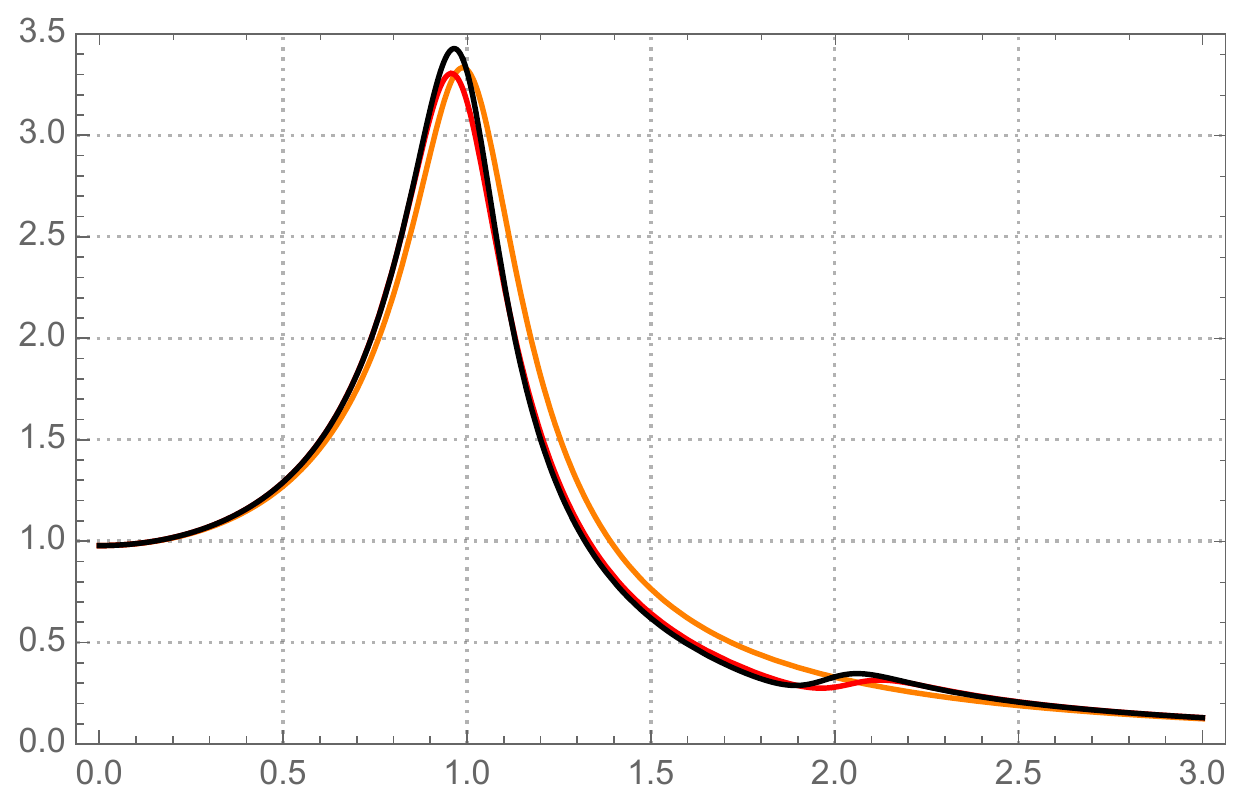}
 \caption{Amplitude of forced $x_1$ versus driving frequency; orange (lightest) is no $k_2$ or $x_2$, red is the exact solution, black (darkest) treats the back reaction to first order in $\alpha$; See section IIIA: 0-0D for explanation of the parameters: $\omega_1 = 1$ and $\omega_2 = 2$, $\alpha = 0.05$ and $\gamma_1=\gamma_2 = 0.3$. }
 \end{figure}
\begin{figure}[h!]
 \includegraphics[width=4.2in]{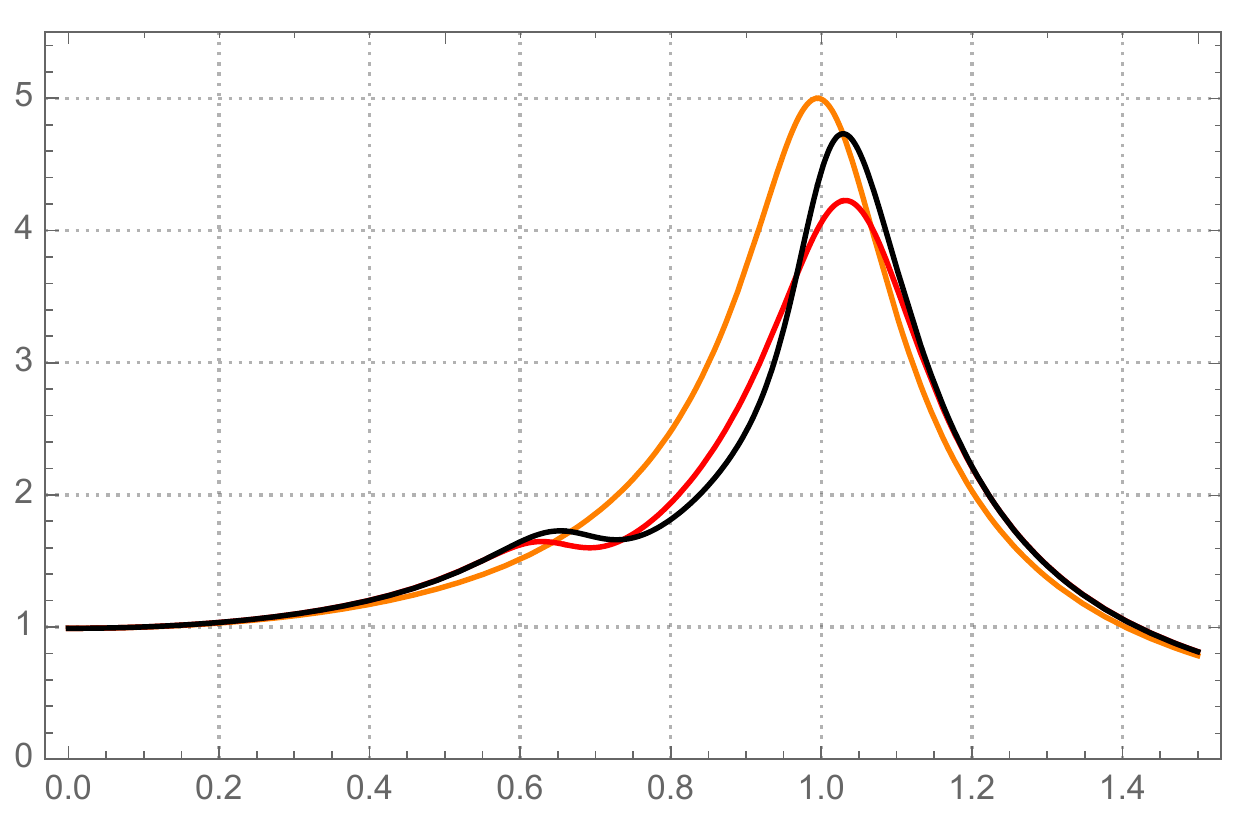}
 \caption{Same as above except $\omega_2 = 2/3$, $\alpha = 0.1$, and $\gamma_j = 0.2$. }
 \end{figure}

\subsubsection{0-1D}

The 0-1D system involves compression waves in the 1D air column.
\medskip

``Linearity" is a key property of the relevant differential equations that allows simple analysis.  It is a defining feature of the ideal 0D spring and a very good approximation for most air motion in musical acoustics.  The 0-1D system is a reminder that one must also consider whether the coupling between two subsystems is approximately linear.  In an abstract 1D wave system, the wave amplitudes and the spatial dimension on which the waves live are unrelated.  Electromagnetic waves are an example.  The wave amplitudes are electric and magnetic fields.  The fields have values at each point in our space.  Magnitudes measured in Tesla and Volts-per-meter cannot be compared to distances in meters.  For the coupling of a 0D to a 1D system of the type described by FIG.~2, the maximum amplitude of the 0D motion at some particular frequency must be small compared to the wavelength of the corresponding 1D wave for the coupling to be linear.  This is satisfied by the physical parameter values relevant to normal playing of a banjo.
\begin{figure}[h!]
 \includegraphics[width=4.3in]{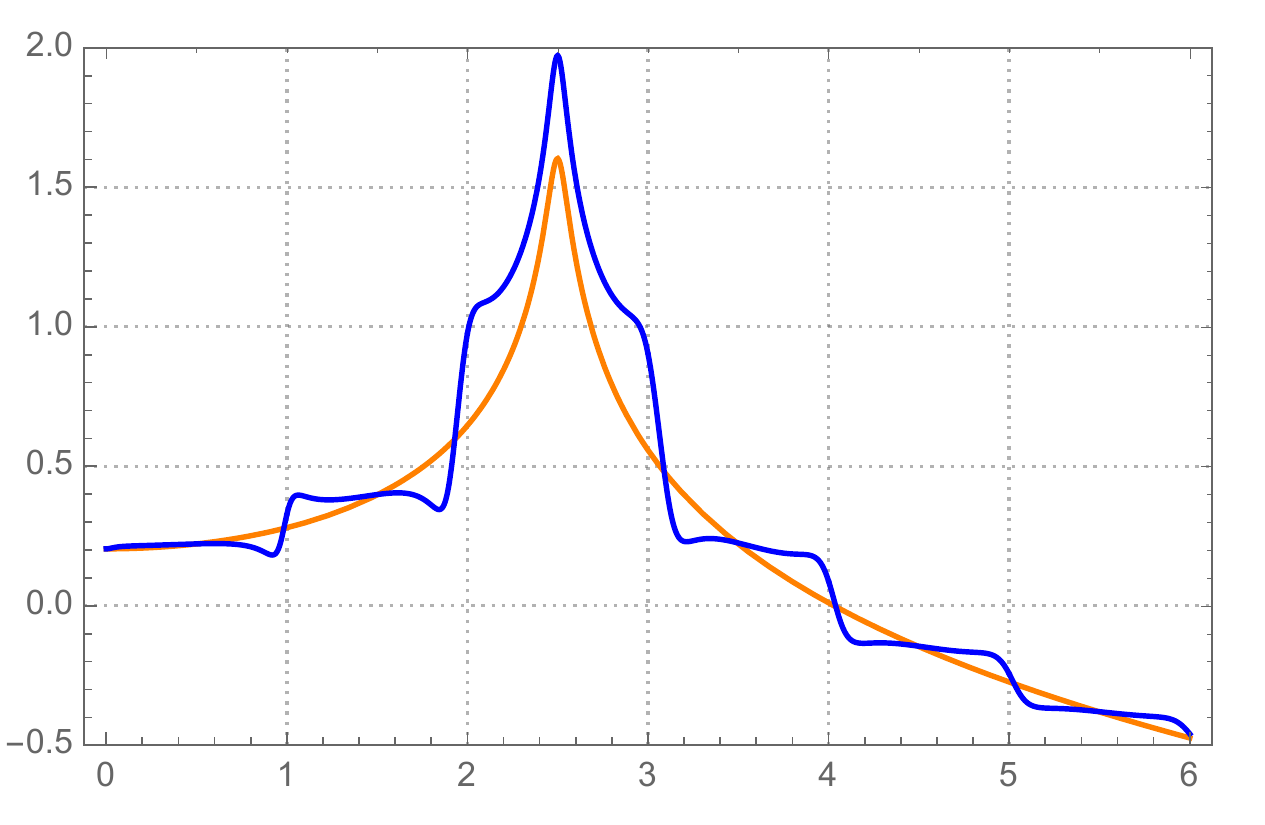}
 \caption{Logarithmic representation of the amplitude of response of piston as a function of driving frequency.  The forced spring by itself is the orange (lighter) curve.  The blue (darker) curve includes the air column back reaction.  $c\rho/m$ = 0.5;  $L/c=\pi$; $\omega_0$ = 2.5; $\gamma_0 = 0.2$; $\gamma_{\text{air}} = \gamma_0(1+0.3\omega)$.  See section IIIB for explanation of the parameters.}
 \end{figure}

The air column resonances appear in the motion of the forced piston.  If one ignores all damping, that back reaction is periodic in frequency.  In particular, each column resonance enhances or alters the uncoupled piston motion by the same factor or multiplicative function.

FIG.~7 compares the piston amplitude as a function of driving frequency with and without the 1D column.  The vertical, amplitude scale is logarithmic.  The scale value and zero are arbitrary.  The message is the shape on a log scale.  See section IIIB for explanation of the parameters.  

\subsubsection{1-2D}

Basic mechanics determines the general structure of the back reaction as a function of frequency.  There are terms with poles (i.e., zeros in the denominator) at the 2D air cavity resonance frequencies.  Each pole residue (numerator) is a measure of the effectiveness of the driven head to excite the air resonance (which depends on the relation of the shapes) and of the effectiveness of the air resonance to drive particular head motions.

To get a definite result for the 1-2D system that is simple enough to be comprehensible, the task is posed in a somewhat simplified manner.  The back reaction strategy works the  same way here as in the previous examples.  An external force, sinusoidal in time, is applied to the 1D ``head."  In general, that force might have a spatial structure that is a superposition of the spatial structures of the modes of the head.  To simplify and sharpen the focus, the explicit start of the calculation is a particular 1D spatial form for zeroth order motion of the head, to be examined as a function of sinusoidal frequency.  This drives the modes of the 2D system.  The back reaction force of that 2D motion on the head is evaluated.  A feature of the 1-2D head-air coupling is that the back reaction includes spatial shapes that are different from the driving force.  An example is illustrated in FIG.~12 in section IIIC.

FIG.~8 is an example of the projection of the back reaction onto the particular driving force's spatial structure.  The vertical scale is a logarithmic measure of the head amplitude of a particular shape for a driving force of that same shape.  In the case illustrated, two half-waves fit on the head.  In the absence of the air, that head mode resonates at frequency $\omega = 4$.  The orange (light) curve is that head resonance by itself.  The blue (dark) curve includes the back reaction of the enclosed air.  The peaks reflect the various air cavity resonances.
\begin{figure}[h!]
 \includegraphics[width=4.3in]{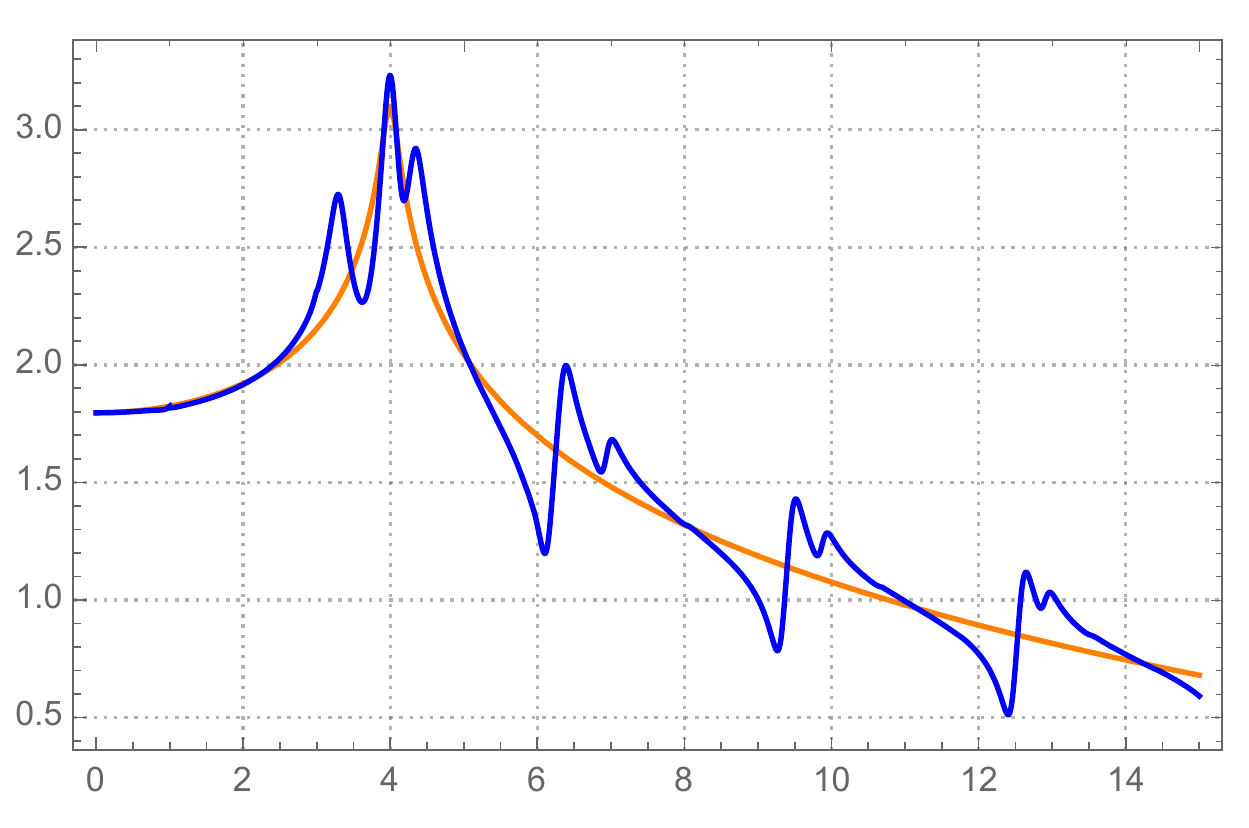}
 \caption{Log of the head amplitude projected onto the two half-wave mode vs $\omega$.  See section IIIC for details.}
 \end{figure}

\subsubsection{the internal resonator as a further example}

The internal resonator is a design element dating back to Fred Bacon in 1906.\cite{internal-resonator}.  Something of a cult favorite to this day, the design creates a somewhat fuller sound.      
 \begin{figure}[h!]
 \includegraphics[width=5.5in]{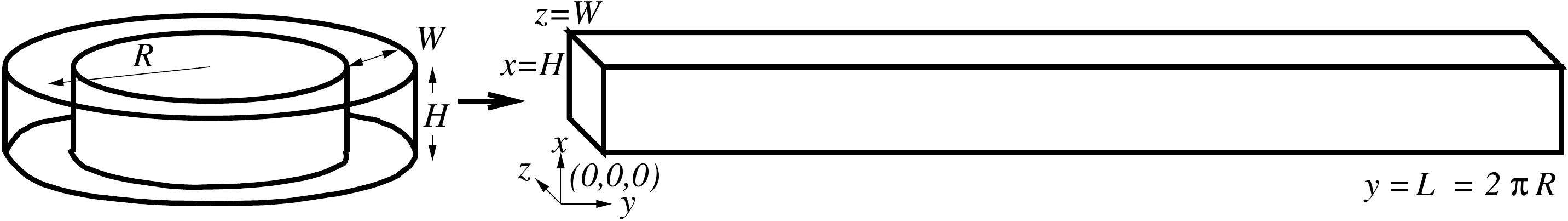}
 \caption{The outer annulus of an internal resonator unrolled into a rectangular box, periodic in $y$}
 \end{figure}
\noindent With a few simplifying approximations, this serves as a straightforward example of extending the 1-2D analysis to a 2-3D system.  The internal resonator gives a back reaction that starts at a much lower frequency than occurs without it.  It also increases the density of pot air resonances as a function of frequency for mid to high frequencies.  FIG.~9 is a sketch of this geometry and indicates the first step in its analysis.

\subsubsection{the Helmholtz resonance}

Finding Helmholtz resonance behavior in the solution of the basic wave equations would be a daunting task because it relies on complicated boundary conditions (i.e., the small mouth or neck that connects to the open air) and involves complicated air motion.  So that is not attempted here.  Rather, employing the standard spring \& mass (0-0D) simplified picture, one can understand how it is excited and pushes back.

\subsubsection{2-3D}

2-3D would be a model of an actual banjo.  No such calculation is attempted here explicitly.  For a circular head and pot, the relevant nodal lines and planes are equally spaced diameters and diametric slices.  Rotational symmetry ensures that the back reaction of the pot air preserves the number of azimuthal nodes.  Hence, the addition of the azimuthal dimension adds computational but no conceptual difficulties.  The radial behavior is an example of a 1D head driving a 2D cavity, but the relevant eigenfunctions are Bessel  functions rather than sines and cosines.  Again, this is a computational complexity rather than a conceptual one.

For a rectangular banjo\cite{mike-gregory}, the obvious rectangular coordinates are also separable.  The motions added by the extra dimension in going from 1-2D to 2-3D are the obvious set of denumerable possibilities.  While the circular banjo has rotational symmetry (and preserved transformation properties under rotations), rectangular boundary conditions break translation symmetry.  That is why the nodal structure of the driving head can be altered by the back reaction --- in contrast to the azimuthal structure in a circular geometry.

\section{equations \& calculations}

The details that follow assume familiarity with forced, damped, coupled oscillators, with  the wave equations for an ideal membrane and for sound and their solutions (including evanescent waves); with Fourier analysis; and with the complex number representation of each of these.  That necessary background is typically included in second-year college physics and will not be reviewed here.  However, one need not know any Bessel function identities, which Hankel function is which, or how to integrate Green's functions over bounded volumes.\cite{greens}

\subsection{0-0D}

The equations of motion for the two oscillators depicted in FIG.~1 are

\centerline{ $m_1 \ddot x_1 = - k_1 x_1 -  k_2( x_1 - x_2) - m_1\gamma_1\dot x_1 + F_{\text{ext}}(t)$}

\centerline{ $m_2 \ddot x_2 =   -k_2(x_2 - x_1)  - m_2\gamma_2\dot x_2$  .}
\medskip

Let
$\omega_1 = \sqrt{k_1/m_1}$, $\omega_2 = \sqrt{k_2/m_2}$, $\alpha = m_2/m_1$, 
$F_{\text{ext}}(t)/m_1 =  {\cal F}$ .  Then
%\bigskip

\centerline{$\ddot x_1 = - (\omega_1^2 + \alpha\omega_2^2)x_1 - \gamma_1 \dot x_1 + \alpha\omega_2^2 x_2 + {\cal F}$}

\centerline{$\ddot x_2 = \omega_2^2 x_1 - \omega_2^2 x_2 - \gamma_2 \dot x_2$  .}
%\bigskip

Let

\centerline{${\cal F}(t) = f  \text{e}^{i\omega t}$}

\centerline{$x_i(t) = {\cal A}_j \text{e}^{i\omega t}$ (with $j=1,2$)  .}

\noindent Note that the ${\cal A}_j$ are complex, i.e., their phases encode the phase shifts relative to $F_{\text{ext}}$ that arise because of the damping.  The external force is applied to oscillator \#1, and the quantity of final interest is $|{\cal A}_1|$.

Use the $\ddot x_2$ equation to find ${\cal A}_2$ in terms of ${\cal A}_1$.  Use that in the $\ddot x_1$ equation to solve for ${\cal A}_1$.  

Let

\centerline{$\Omega_j^2 = \omega^2 - i \gamma_j \omega$ (with $j=1,2$) .}

Then

\centerline{ ${\cal A}_1 = f / \{\omega_1^2 - \Omega_1^2 - \alpha [{{\omega_2^2 \Omega_2^2 \over {\omega_2^2 - \Omega_2^2}}}]\}$  .}
\medskip

That is the exact solution.  The back-reaction calculation begins with the solution of the \#1 motion in the absence of coupling to \#2.  That means taking $k_2=0$ in the equations of motion, which amounts to setting $\alpha = 0$ in the $x_1$ amplitude:

\centerline{${\cal A}_1^{(0)} = f / \{\omega_1^2 - \Omega_1^2\}$  .}

Use that $x_1^{(0)}(t)$ solution, i.e., with ${\cal A}_1^{(0)}$ in the $\ddot x_2$ equation, to find $x_2^{(0)}(t)$.

Use $x_2^{(0)}(t)$ to find $x_1^{(1)}(t)$, the solution of 

\centerline{ $m_1 \ddot x_1^{(1)} = - k_1 x_1^{(1)} -  k_2( x_1^{(1)} - x_2^{(0)}) - m_1\gamma_1\dot x_1^{(1)}$ .}
%\bigskip

\noindent In particular, the motion $x_2^{(0)}(t)$ produces the force that drives $x_1^{(1)}(t)$.

The total motion of \#1, including the first order back reaction, is $x_1^{(0+1)}(t)$ which equals  $x_1^{(0)}(t) + x_1^{(1)}(t)$.  The (complex) amplitude of $x_1^{(0+1)}(t)$ is

\centerline{${\cal A}_1^{(0+1)} = {f \over {(\omega_1^2 - \Omega_1^2)}} \{1 + \alpha {{\omega_1^2 \Omega_2^2} \over {(\omega_1^2 - \Omega_1^2)(\omega_2^2 - \Omega_2^2)}}\}$  .}
\medskip

If the exact ${\cal A}_1$ is expanded in powers of $\alpha$, the sum of the zeroth and first order terms in $\alpha$ is precisely ${\cal A}_1^{(0+1)}$.  Higher terms in the expansion are clearly iterations of back reaction.  Furthermore, this procedure is directly applicable to higher dimensional extended systems as long as the systems and their coupling can be treated as linear and the interaction between the two systems is weak, e.g., because of a large impedance mis-match.

Three versions of $|{\cal A}|_1/f$ are plotted in FIG.s 5 and 6 versus $\omega$ for some suggestive parameter values.  The three versions are (in orange [lightest]) the uncoupled single forced damped oscillator; the exact solution (in red) of the coupled system; and the lowest order back-reaction corrected solution (in black).  The horizontal axis is frequency; $\omega_1 = 1$ in each case; and $\omega_2$ is higher in FIG.~5 and lower in FIG.~6, as indicated. Values of the small coupling parameter $\alpha$ ($ = m_2/m_1$) are given;   the damping constants $\gamma_j$ are taken to be equal, with values as indicated.

\subsubsection{\bf a comment on the role of damping}

As noted above, the amplitude of the steady-state response of a single forced damped oscillator is

\centerline{${\cal A} = {f \over {\omega_0^2 - \omega_0^2 +i \gamma \omega}}$  .}

 \begin{figure}[h!]
 \includegraphics[width=3.0in]{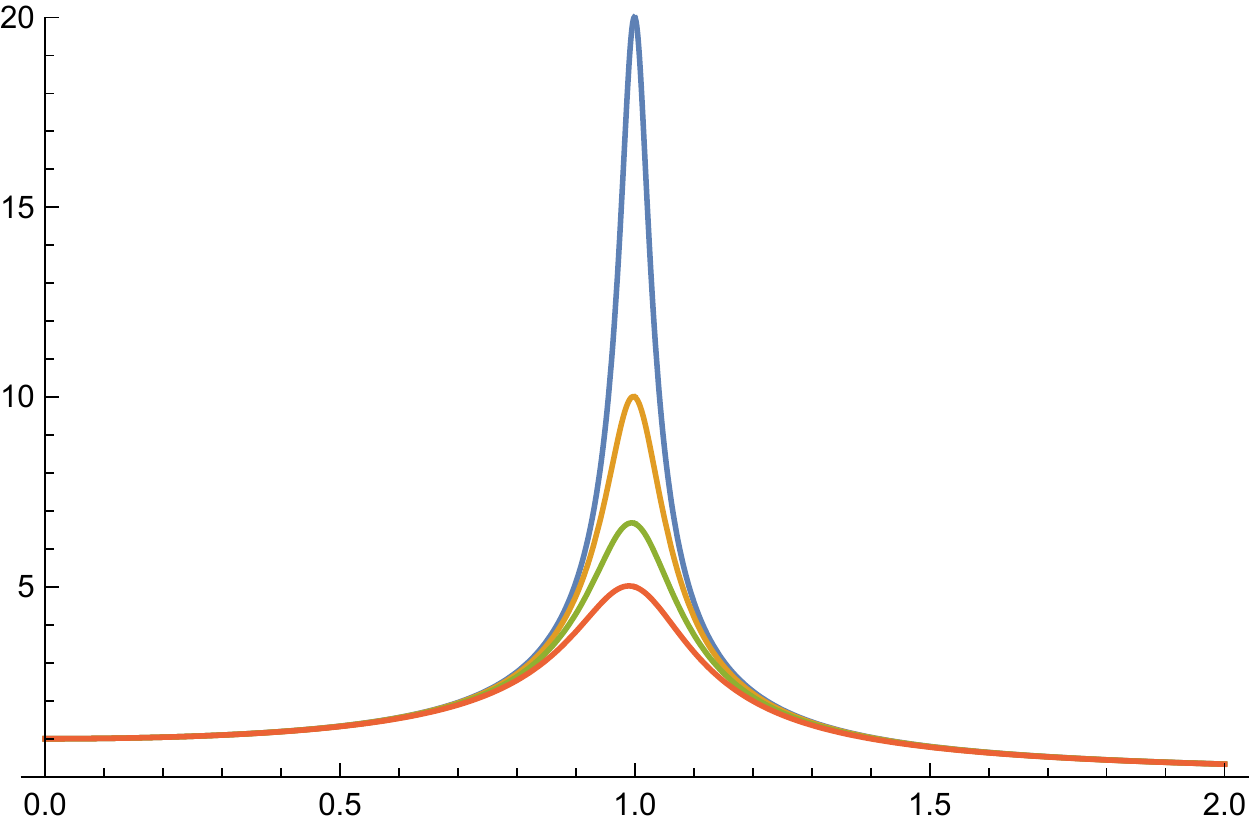}
 \caption{amplitude of a forced damped oscillator with successively larger damping, i.e. $\times1$, $\times2$, $\times3$, and $\times4$, and other factors unchanged}
 \end{figure}

It is said that increasing $\gamma$ makes the resonance broader.  However, it is only broader relative to its maximum, e.g. ``full width at half max."  With the other parameters held fixed, increasing $\gamma$ just makes the resonance weaker, as illustrated in FIG.~10.

For zero damping, the response is in phase with the sinusoidal driving force --- up to a sign or $180^{\text{o}}$.  It is in phase below and $180^{\text{o}}$ out of phase above; i.e., ${\cal A}$ is real, positive below $\omega_0$, and negative above; and $|{\cal A}|\to \infty$ as $\omega \to \omega_0$.

With $\gamma \neq 0$, ${\cal A}$ is complex.  At resonance it is imaginary.  Consequently, the positive definite $|{\cal A}|$ is also large at resonance.  If such a resonance is providing some sort of back reaction to some zeroth order amplitude, away from the resonance it contributes with opposite signs.  Near the resonance, it adds ``in quadrature" to the uncoupled forced motion, i.e., as the square-root of the sum of the squares, to produce the total amplitude, because of the $90^{\text{o}}$ phase difference.

\subsection{0-1D}

The 0-1D system is a spring loaded, driven piston in a (one-dimensional) cylinder.  The new aspect is the linear, dissipationless compression waves supported by the ``air" in the cylinder. The illustration in FIG.~2 defines the signs of the coordinates.  The wave equations are

\centerline{$\dot p = - c^2 \rho v'$}
\centerline{$\rho \dot v = - p'$}
\bigskip

\noindent where $p(z,t)$ is the pressure, $v(z,t)$ is the air velocity, $c$ is the speed of sound, and $\rho$ is the density of the air.  Because the system is one-dimensional, $p$ is a force, and $\rho$ is the mass per unit length.  (Alternatively, one may think of it as a three-dimensional cylinder with unit cross section.)

These equations can be combined to yield the wave equations for $p$ and $v$ separately:

\centerline{$\ddot v = c^2 v''$}
\centerline{$\ddot p = c^2 p''$}
%\bigskip

The math is simple only in the absence of any dissipation.  So that is how we proceed at this point.  Ultimately, dissipation will prove crucial to the form of the final behavior, but it is not only difficult to model analytically, it has many different physical origins with very different forms and magnitudes.  So a discussion is postponed until later.

In the absence of dissipation, all motion is in phase (up to a sign) with the driving force, and there is no need (as yet) to use complex numbers to represent the possible phases.  For a sinusoidal force $F(t)$ applied to the piston (of mass $m$ on a spring of spring constant $k$ and $\omega_0 ^2 = k/m)$, the spring amplitude $\psi(t)$ is in phase with $F$ and $\dot \psi(t)$ is $90^{\text{o}}$ out of phase.  The piston--air interaction is modeled by the constraint that $v(L,t) = - \dot \psi(t)$, where $L$ is the length of the cylinder, and the piston is at the $z=L$ end.  (The ``--" comes from the sign convention of FIG.~2.)  

The closed end has $v(0,t) = 0$, while the piston end must have the form

\centerline{$v(L,t) = A \;\text{sin} \,\omega t$ .} 
%\bigskip

Writing $v(z,t)$ as $g(z)\;\text{sin}\,\omega t$, $g(0) = 0$, $g(L) = A$ and

\centerline{$g''(z) = - {\omega^2 \over c^2} g(z)$ .}  

Hence,

\centerline{$g(z) = B \;\text{sin} \,{\omega \over c} z$}
%\bigskip

and

\centerline{$B = A/\text{sin}\,{{\omega \over c} L}$  .}
%\bigskip

This result for $v(t)$ in the equation for $\dot p$ implies 

\centerline{$p(L,t) = c \rho A \;\text{cos}\,\omega t / \text{tan} \,{{\omega L}\over c} $ .}
%\bigskip

$p(L,t)$ is the back reaction force on the piston if the amplitude of $v(L,t)$ is given by $A$.  The chosen sign convention is that a positive $p$ is a force on $\psi$ in the negative $\psi$ direction.  $A$ is determined by matching to the zeroth order motion $\psi^{(0)}(t)$ of the externally forced piston.  $\psi^{(0)}(t)$ satisfies the equation

\centerline{$m \ddot \psi^{(0)} = - k \psi^{(0)} + F$}
%\bigskip

\noindent where $F(t) = m f \text{cos} \omega t$ .  The solution is

\centerline{$\psi^{(0)}(t) = {\cal A}^{(0)}(\omega)\, \text{cos}\,\omega t$}
%\bigskip

\noindent and

\centerline{${\cal A}^{(0)} = f / (\omega_{0}^2 - \omega^2)$}
%\bigskip

\noindent with $\omega_0^2 = k/m$.  That implies that

\centerline{$A^{(0)} = - {{f \omega} \over {\omega_0^2 - \omega^2}}$  .}
%\bigskip

\noindent So the back-reaction pressure (or force) is

\centerline{$p^{(0)}(L,t) = - c \rho {{f\omega} \over {\omega_0^2 - \omega^2}} {{\text{cos}\,\omega t} \over {\text{tan}{{\omega L}\over c}}}$ .}
\medskip

This is the force that drives $\psi^{(1)}$.  $\psi^{(0)}$ and $\psi^{(1)}$ are added (superposed) to form $\psi^{(0+1)}$, whose amplitude is

\centerline{${\cal A}^{(0+1)} = {f \over {\omega_0^2 - \omega^2}}[1 + {{c\rho} \over m}{\omega \over {(\omega_0^2 - \omega^2) \text{tan}{{\omega L} \over c}}}]$  .}
\medskip

The zeros of tan${{\omega L} \over c}$ are the resonances of the length $L$ column, closed at both ends.  The coefficient {\large ${{c\rho}\over m}$} represents the impedance mis-match between the heavy, stiff piston and the much more easily compressible air. However, it is not dimensionless.  It has the units of 1/time.  Its size depends on the units used to measure it.  Formally, the back-reaction is an expansion in powers of this coefficient. For that to be a useful expansion, the correction to the ``1'' term must be small --- and successive terms smaller still.  So the dimensionless parameter must have a further factor of frequency in its denominator.

Damping tames the poles of 1/tan${{\omega L} \over c}$ so that what remains is ${\cal O}(1)$.  {\large${1\over {\omega_0^2 - \omega^2}}$} away from its pole is ${\cal O}(1/\omega_0^2)$ or ${\cal O}(1/\omega^2)$, whichever is larger.  So {\large${c \over {\omega L}}$} must be a long time compared to the other relevant times that characterize the motion, with ``relevant'' depending on what is being asked. 

The result can be described qualitatively as the obvious: the closed pipe resonances enhance the piston motion when the driven piston excites them.  The purpose of this whole exercise is to confirm that the obvious notion is, indeed, correct physics.

Constructing a useful plot to illustrate the qualitative features requires dealing with damping.  At this point, an engineering, practical approach is appropriate.  The substitution $\omega_0^2 - \omega^2 \to \omega_0^2 - \omega^2 + i\gamma_0 \omega$ is a commonly used representation of damping of the spring.  Dissipation of the air motion is definitely not dominated by viscosity, at least in the context of music and sound.  The dominant effects are losses to the bounding materials that absorb more energy than they return.  Examples include motion of the side walls, the head itself, and sound radiation.  This most significantly alters the no-dissipation formulas in the vicinity of the zeros of tan${{\omega L} \over c}$.  A reasonable model is the replacement within the tangent function $\omega \to \omega -i\gamma_{\text{air}}/2$, where $\gamma_{\text{air}}$ is some small, real parameter.  A single $\gamma_{\text{air}}$ would represent an equal decay time for all of the air resonances.  In practice, the higher frequency resonances typically decay faster.  So  $\gamma_{\text{air}}$ could be chosen to be an increasing function of $\omega$.

FIG.~7 gives a logarithmic representation of the amplitude of response of piston as a function of driving frequency.  ``Logarithmic" makes it akin to some sort of decibel plot. The forced spring by itself is the orange curve.  The (darker) blue curve includes the air column back reaction.  For this particular plot, the parameter values are: $c\rho/m$ = 0.5;  $L/c=\pi$; $\omega_0$ = 2.5; $\gamma_0 = 0.2$; $\gamma_{\text{air}} = \gamma_0(1+0.3\omega)$.  The air column resonances appear at integer values of the frequency, while the uncoupled piston resonance is at 2.5.  The damping ``constant" of the air resonances matches the piston's value, $\gamma_0$, as $\omega \to 0$ and increases linearly, e.g., going through 1.9$\gamma_0$ at $\omega=3$.  The air resonance enhancement is greatest for resonant frequencies nearest the original piston frequency.  And it decreases with increasing frequency as the damping of those air resonances increases.

\subsection{1-2D}

If the head is a one-dimensional membrane (i.e., a string) and the pot is two-dimensional,  new, challenging features arise.  Driving the head with an external force requires specifying the spatial structure of the drive as well as its time dependence.  And the 2D air motion is considerably more complicated.  The pressure, $p({\bf r},t)$ is still a scalar, but the air velocity, ${\bf v}({\bf r},t)$ is a two-vector.  Separation of variables for the relevant differential equations in simple geometries aids the analysis.  However, there is a fundamental mismatch between the vibrating boundary (the 1D head) and the enclosed (2D) air.  This is related, in part, to two conflicting considerations.  The head motion pushing in some direction generally results in some air motion in an orthogonal direction.  But the boundary conditions include sides where the air motion must vanish.  In fact, the motion resulting from the simplest possible forcing can only be described as an infinite sum of simple motions.

In the 0-1D example, the external, driving force was taken to be sinusoidal in time --- not because most or many forces actually are sinusoidal but because any time dependence can be written as a sum of sinusoids of different frequencies.  Each one produces a characterizable and calculable response.  And, if the system is linear, the total response is the sum of the various sinusoidal responses.  For the 1D head of the 1-2D system, we use the same strategy, consider some normal mode shape of the 1D system, and study the response as a function of frequency.  In particular, we consider a force on the head whose spatial structure produces a particular sinusoidal displacement of the head with the force's frequency.  That moves the adjoining air at that frequency.  The goal is to understand the back reaction pressure, which gives a small additional force on that head motion.  However, there is nothing simple about it.  Off-resonant forcing of an enclosed compressible fluid is complicated.  Choosing a particular sinusoidal boundary vibration is only a partial simplification of the general problem.   

As depicted in FIG.~3, the head equilibrium position is along $x=H$ and stretches in the $y$-direction from 0 to $L$.  Head motion is represented by $\psi(y,t)$, and $\psi > 0$ is deformation in the +$x$-direction.  $\psi$ satisfies

\centerline{$\ddot \psi = c_s^2 \, \psi'' + F(y,t)/\rho_s$}
%\bigskip

\noindent where $c_s$ is the wave speed of $\psi$ along the head, $F(y,t)$ is the externally applied force, and $\rho_s$ is the mass per unit length of the head.

For the particular $F$

\centerline{$F(y,t) = \rho_s f \;\text{sin}\,ky \;\text{cos}\,\omega t$ ,}

\noindent the steady state $\psi$ takes the form

\centerline{$\psi(y,t) = A \; \text{sin}\,ky \;\text{cos}\,\omega t$}

\noindent with

\centerline{$A = {f \over {c_s^2 k^2 - \omega^2}}$  .}

\noindent While $\omega$ can take any value between 0 and $\infty$, $k$ must produce $\psi$ solutions that fit the fixed-end boundary conditions.  Therefore $k L / \pi$ must equal some integer $n$, which is the number of half-waves in $L$.

The above is the form taken for $\psi^{(0)}$, which serves as a boundary condition for the air:

\centerline{$v_x(H,y,t) = \dot \psi^{(0)}(y,t)$ .}
%\bigskip

The wave equations for {\bf v} and $p$ are

\centerline{$\dot p = - c^2 \rho \nabla \cdot {\bf v} $}
\centerline{$\rho \dot {\bf v} = - \nabla p$}

\noindent which give

\centerline{$\ddot p = c^2 \nabla^2 p$}
\centerline{$\ddot {\bf v} = c^2 \nabla^2 {\bf v}$}

\noindent where $c$ is the speed of sound in the 2D system and $\rho$ is its mass per unit area.

With rectangular boundaries, these equations are separable in Cartesian coordinates.  There is a complete set of solutions of the form

\centerline{$p(x,y,t) =  p_x(x)\: p_y(y)\: p_t(t)$}

\noindent with the $p_i$, $i=x, y, t$, each satisfying its own harmonic/Helmholtz-like equation, whose solutions are sinusoidal.  ``Complete set" means that any particular solution can be expressed as a superposition of the various factored or ``separated" solutions.  To match the sinusoidally forced boundary, a superposition of an infinite series will be required.

%\bigskip

We again seek a persistent solution with sinusoidal frequency $\omega$.  For a given set of boundary conditions, the solution is unique, and it is finite away from the enclosure resonances.  In the limit that damping can be ignored, the pressure and velocity at each point vary (in time) in phase or $180^{\text{o}}$ out of phase (just a -- sign) with the driving.  The solution is made up of the standing wave superpositions of traveling waves which have wave vectors {\bf k}.  The {\bf k}'s satisfy $\omega^2 = k_x^2 + k_y^2$.  So, for a given $\omega$ and $|k_y|$, there are four distinct {\bf k}'s.  In general, no single value of $k_y$ will lead to a superposition of waves that satisfy the boundary conditions.  How much of which {\bf k}'s are needed will depend on the exact nature of the forcing.

The spatial mode shapes of the 1D head are a natural basis for the boundary condition forcing of the 2D cavity.  An infinite series of air motion $k_y$'s will be required to match a single $k$ of the head.  However, the motion corresponding to each  term in that series can be computed using simple math.

The $x=H$ boundary is forced at $\omega$ as shown in FIG.~3 with a wave number $k_\psi$ to give
\centerline{$\psi^{(0)}(y,t) = A \;\text{sin}\,k_\psi y \;\text{cos}\,\omega t$ .}
\noindent So we seek solutions for the air motion for which

\centerline{$v_x(H,y,t) = \dot \psi^{(0)}(y,t)$ }
\noindent and $v_x(0,y,t) = 0$ and $v_y(x,0,t) = v_y(x,L,t) = 0$.
%\bigskip

No simple solution exists, i.e., with $k_y = \pm k_\psi$ and $k_x = \pm \sqrt{{\omega^2 \over c^2} - k_y^2}$.  Rather, a solution is provided by expanding that $\text{sin}\,k_\psi y$ as a cosine Fourier series in $k_y$. (!)  Each individual term in the series, when used as a boundary condition on $v_x(H,y,t)$, gives a simple solution for {\bf v} and $p$.  The desired solution is the infinite sum of the whole series.  A truncation of that series gives an approximation to the solution.

Often glossed over in teaching and learning Fourier series is an ambiguity in the  representation of a function on a finite interval.  Different Fourier series representations that converge to the same function within the interval may have different forms when extended beyond.    In physics applications, the physics usually dictates which is correct or most convenient.  
 \begin{figure}[h!]
 \includegraphics[width=2.5in]{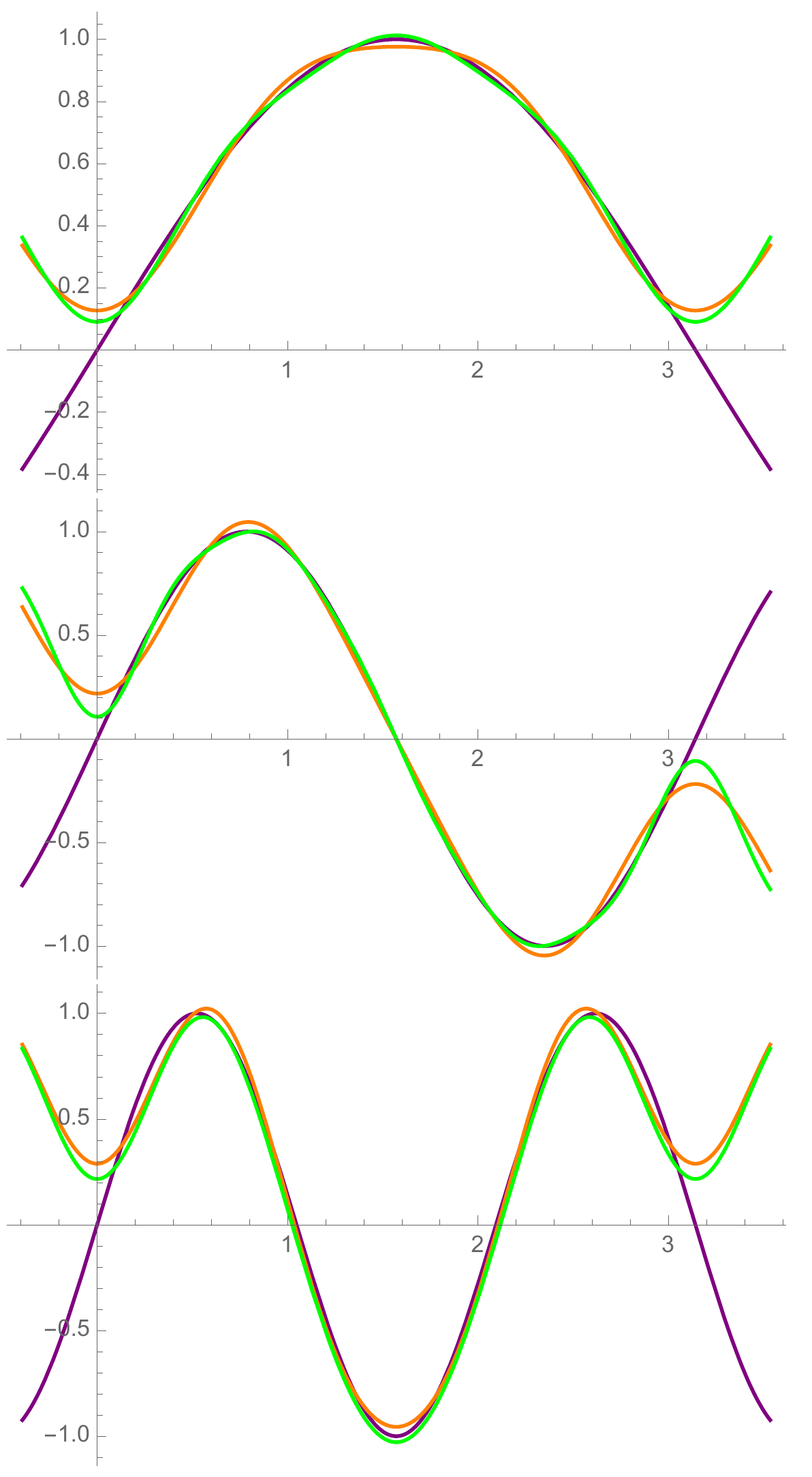}
 \caption{The cosine series in the graphs (top to bottom) have 3 and 4 terms,  3 and 6 terms, and 4 and 8 terms, respectively }
 \end{figure}

Three pictures are worth at least three thousand words.  In the three graphs of FIG.~11, the interval of physical interest is 0$\,<\,$$x$$\,<\,$$\pi$.  The functions in that interval are sin\,$x$, sin\,$2x$, and sin\,$3x$.  Each can be represented as the obvious: a single term of a Fourier series in sines.  Those are the purple (dark) curves, and the functions' values are shown a small bit beyond the region of physical interest.  A Fourier series in cosines is easy to construct whose sum represents the same function on 0 to $\pi$ in alternating intervals of length $\pi$ but has the opposite sign in the intervening intervals of length $\pi$.  For the initial sine functions of interest, this construction yields a function that is continuous with a discontinuous first derivative at multiples of $\pi$.  That means that the coefficients for large $j$ go like $1/j^2$.  The relevant particular example is the following:

\centerline{$f(x,n) = \text{sin} \,xn = \sum_{j=0}^\infty \: a_j (n) \:\text{cos}\, xj$ ;}

\centerline{if $n+j$ is odd, $a_j(n) = {4 \over \pi}{n \over {n^2 - j^2}}$ if $j\ne 0$ and $a_0(n) = {2 \over \pi n}$ if $j=0$ }

\centerline{ and $a_j(n) = 0$ if $n+j$ is even.}

\noindent For a given $n$, the largest coefficients are for $j=n \pm 1$, even though the $j=n$ coefficient vanishes.  This is the mathematical reflection of how the wavelength of the head motion imprints itself on the air motion.

So we seek a persistent solution of the air equations with 

\centerline{$v_{x,j}(H,y,t) = B_j \,\text{cos}\,k_{y,j} y \;\text{sin} \,\omega t$ }

\noindent with $k_{y,j}  = \pi j/L$ and $j = 0,1,2,3...$ and $  k^2_{x,j} = \omega^2/c^2 - k^2_{y,j}$.  Until it is time to exhibit the $j$-dependence and the sum over $j$, the $j$ labels will be suppressed.

For any particular $j$, with $D = B /( k_x \;\text{sin}\, k_xH)$, all boundary conditions and wave equations are satisfied by

\centerline{$v_x(x,y,t) = k_x D\;\text{sin}\,k_x x \;\text{cos}\, k_y y \;\text{sin} \,\omega t$}

\centerline{$v_y(x,y,t) = k_y D \; \text{cos}\,k_x x \; \text{sin}\, k_y y \; \text{sin}\, \omega t$}

\centerline{$\  \  \  \ p(x,y,t) = \rho \omega D\; \text{cos}\,k_x x \;\text{cos} \,k_y y \;\text{cos}\, \omega t$ .}

For $j >  {{L \omega} \over {\pi c}}$, $k_{x,j}$ is imaginary.  Nevertheless, the solution, when properly interpreted, is precisely as given.  Simply replace the sine and/or cosine of the imaginary angle with their hyperbolic counterparts.  The hyperbolic functions are, in fact, trig functions of imaginary arguments.  The boundary conditions on {\bf v} are still respected.  

%\bigskip
\bigskip

\centerline{\bf THE 1-2D INFINITE SERIES SOLUTION:}

\medskip

Collecting all the pieces so far, we can construct the solution for the first order back reaction of the air on the head due to a particular motion of the head.  If the zeroth order  head motion (i.e., ignoring the air) is 

\centerline{$\psi^{(0)}(y,t) = A \; \text{sin} \, {\pi \over L}ny \; \text{cos} \, \omega t$ ,}

\noindent then the back reaction force per unit length along the head is the air pressure at $x=H$:
\medskip

\centerline{$p(H,y,t) = - \rho \omega^2 A \, \sum_{j=0}^\infty \:$$ {\large {a_j(n) \over {k_{x,j}{\text{\small tan}}\,k_{x,j}H}}}$$ \; \text{cos} \, k_{y,j}y \; \text{cos} \, \omega t$}
\medskip

\noindent with $k_{y,j} = {\pi \over L}j$ ,  $k_{x,j} = \sqrt{{\omega^2 \over c^2} - k_{y,j}^2}$ , and $a_j(n)$ as given above.

\medskip

As simplified and idealized as the 1-2D system is, its behavior captures several features that are operative in a real banjo.  By construction, in this calculation the pressure on the head has the time dependence of the original sinusoidal forcing.  If we ignore the air, the external force applied to the head would only excite one head mode.  The amplitude of response would depend on how close the force's frequency is to that mode's resonance.  For a given sinusoidal $y$ dependence, there is an infinite series of air normal modes that differ from each other by their $x$ dependence and have the corresponding resonant frequencies.  However, the nature of the head-to-air coupling mixes different $y$ behaviors.  The back reaction pressure introduces new spatial structure not present in the original $\psi^{(0)}(y,t)$, which has a wavelength in $y$ of $2L/n$.  (It is clear that the cosine series does not re-sum to a single sine function because its coefficients are not simply $a_j(n)$ but have additional strong $j$ dependence.)  This kind of mixing contributes to the musicality of the instrument.  In the end, it is desirable that the responses to different sorts of plucks on the strings vary smoothly with the fundamental frequencies.   If isolated frequencies jump out in the produced sound, it is a bad design.  Such isolated resonances may be features of the various sub-systems, but the coupling between subsystems mixes things up so that the timber of the produced sound is a smooth function of pitch.

On the other hand, $y$ wave numbers near that of the driving force are emphasized by the form of $a_j(n)$.  While there is no $j=n$ term in the sum over $j$, the largest values of $a_j(n)$ are at $j = n\pm 1$ and decrease as the $j$-$n$ difference increases.  

The pressure gets large when any of the $j$-dependent factors in the denominator get small.  The zeroes of the denominator arise because the inevitable damping has been ignored in the derivation.  $k_{x,j} \to 0$ when the driving frequency matches air waves whose $y$ dependence fits in $L$ and are constant in $x$. The wave with $j=n$ is excluded from the list, but there are infinitely many others as we sweep through $\omega$. This is one important feature of systems with higher spatial dimension than the 0-1D model.  Motion of the head in the $x$ direction creates pressure variations which create motion of the air in the orthogonal $y$ direction (or the two directions orthogonal to $x$ in a 2-3D system) and can excite resonances in that direction (or plane).

tan$\,k_{x,j}H \to 0$ when the wave vector component $k_{x,j} = \sqrt{(\omega/c)^2 - (\pi j/L)^2}$ corresponds to fitting the $x$-component of waves in  $H$ and the $y$-component in $L$.  The 2D waves in the sum for which tan$\,k_{x,j}H \to 0$ are all of the normal modes of the closed $L \times H$ box for which $j+n$ is odd. When  $k_{x,j}$ becomes imaginary, all of the sinusoidal trig functions of imaginary arguments become hyperbolic functions.  Those functions satisfy the same differential equations (as is obvious from their complex representations).  They also satisfy the requisite boundary conditions.  So, for $(\omega/c)^2 - (\pi j/L)^2 < 0$, the tan becomes tanh.  The only zero of tanh${\,\theta}$ is at $\theta = 0$, and it is small only in that vicinity.  That behavior is the second half of the account of the $x$-direction resonances associated with $k_{x,j} \to 0$ from the real side.  Away from those $k_x$ values, tanh$\,k_{x,j}H$ is just close to 1.

In summary, the poles of the (undamped) back reaction are at all the closed box resonances and their strength in the sum over resonances is given by their spatial overlap at the driven boundary, as reflected in the coefficient $a_j(n)$.  

\medskip

\centerline{\bf PRESSURE PROJECTED ONTO THE ORIGINAL WAVE NUMBER:}

The back reaction, $p(H,y,t)$, is also a function of $\omega$, $L$, and $k_\psi$.  Its important features were highlighted above.  Those feature's relative sizes and impact depend on all the parameters.  Furthermore, how it all works out depends very significantly on the damping.  A particular projection of the result is of possible interest and is simpler to contemplate and simpler to plot.  The back reaction can be re-expanded as a series of sines on the interval $0<y<L$.  That gives a double Fourier sum.  However, any single sine term will be expressed as a single sum over $j$.  The biggest term is the original, chosen spatial structure of the driving force, i.e., sin$\,k_\psi y$.  That is the back reaction onto the original forcing.  Also of interest is how strongly the back reaction acts on different $y$ wave numbers.

Define such a projection of $p$ along $x=H$ onto a sine with $m$ half-waves as

\centerline{$p(m,...) \equiv {1 \over {2L}}\int_0^L \text{sin}\,{\pi \over L}my \;p(y,...) \, dy$}% {\Large \bf WHY THE 2???}}
\medskip

\hspace{2.43in} $ = -\rho \omega^2 A \sum_{j=0}^\infty {{a_j(m) a_j(n)} \over {k_{x,j} \text{tan}\,k_{x,j}H}}\;\text{cos}\,\omega t$ .
\medskip

\noindent The comments made previously about how each of the terms in the formula can make it large apply here as well.  Of particular importance are the wave numbers that fit in the $L \times H$ box and the near matching of $j$ in the sum to $n$, the number of half-waves in the driving force, and now also to $m$, the $y$ shape projection onto the head.  The $a_j(m) a_j(n)$ factor is largest for $m=n$ and $j = n \pm$ 1.  More generally, $n+m$ must be even, and $n+j$ must be odd.  And the $a_j(m) a_j(n)$ factor gets smaller as the differences of these integers gets larger.

The purpose of a plot of $p(m,...)$  is simply to illustrate some of its qualitative features.  Any such plot requires assignment of numerical values.  However, none of these nor even the model itself realistically represent a musical instrument.  

The back reaction produces poles.  Their locations are of physical origin.  However, as-yet-unmodeled physics must keep the amplitudes not only finite but of modest maximum size.  This was done properly in the basic equations for the 0-0D model and heuristically for the 0-1D model.  Small imaginary parts are added to the resonant frequencies to represent damping, and there are no longer any infinities.   Adding a small imaginary part to the driving frequency represents  different physics but has a similar result.  That describes a driving force that oscillates with some frequency but whose amplitude dies away with a slow exponential in time.  Even if the energy dissipation in the driven system were arbitrarily weak, it would take infinitely long for the driving force to generate an infinite amplitude response on resonance.  If the force dies away before $t = \infty$, then the amplitudes remain finite.

The small imaginary parts of the frequencies are connected to yet one more important aspect of the forced motions.  The phase between the driver and the response cannot change abruptly from $0^{\text{o}}$ to $180^{\text{o}}$.  The range in frequency over which this transition actually occurs is set by the sizes of those imaginary parts.  The complex number representation is just a convenient way to keep track of phase relations among trigonometric functions.

\begin{figure}[h!]
\includegraphics[width=4.7in]{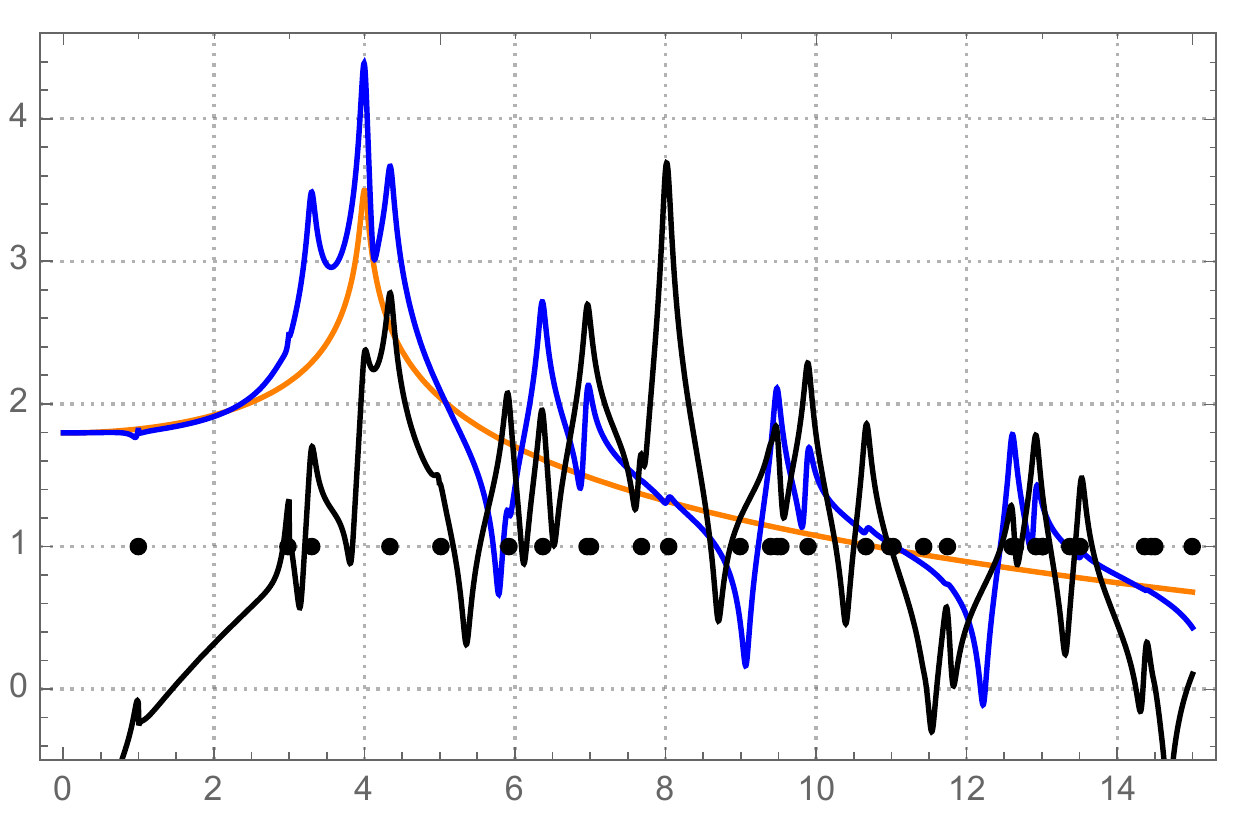}
 \caption{With the external force distributed as $n=2$, the amplitude of the $n=2$ head mode by itself is in orange; including the back reaction as projected onto $n=2$ in blue; and the $n=4$ response to the same $n=2$ driving in black --- all versus the driving frequency $\omega$.  Dots mark the contributing, odd $j$ $L \times H$ ($= \pi \times 1$) air cavity resonant frequencies.  The head resonant frequencies for $n=2$ and $4$  are at $\omega = 4$ and $8$.} 
 \end{figure}
 The 1-2D system presents a further problem regarding what to plot, given that there are so many variables of interest.   FIG.~8 shows the effect of the air pressure back onto the $n=2$ spatial structure that characterized the external forcing. \noindent That result is reproduced in FIG.~12. The zeroth order forced head amplitude is in orange (lightest).  The head motion of $n=2$ including the first back reaction is the blue curve.  The peaks are sharper than in FIG.~8 because the chosen damping constant is smaller.  The underlying zero-damping formula is the same.  (The small parameter that characterizes the impedance mis-match is also a bit smaller than used for FIG.~8.) 

The black curve in FIG.~12 is the amplitude of the $n=4$ motion that results from the $n=2$ driving.  All of the other parameters are exactly the same as for the $n=2$ motion.  Hence, it is legitimately plotted on the same graph with the same scales.  The $n=2$ pure head resonance is at $\omega = 4$.  The $n=4$ head resonance is at $\omega=8$.  The black curve shows clearly that driving the head with the $n=2$ spatial structure gets a strong response at $\omega=8$ from the $n=4$ head mode.

The black dots mark the resonant frequencies of the $L \times H$ air cavity (with $L=\pi$ and $H=1$ and $c=1$) with non-zero coefficients in the sums.  The damping parameter was chosen to emphasize that the back-reaction peaks really do show up at the air resonance frequencies.  That is obscured by stronger damping, when the resonances overlap to a greater extent.

\subsection{the internal resonator}

For an internal resonator banjo (circular, 2-3D), the radial modes of the head do not match simply onto the modes of the air in the pot.  The idealized pot is two separate cavities: an outer annulus and an internal cylinder.  The bottom of the cylinder has an open back.  The bottom of the annulus is closed.  On a real banjo, the internal wall comes up to within about $5/8''$ of the head.  Communication between the two cavities is essential.  That cannot be modeled with simple sums of sines and cosines, no matter how simplified is the rest geometry.  That connection is not {\it a priori} weak or an impedance mismatch.   Experimentally, in going from sealed to a $5/8''$ opening, the air spectrum changes smoothly and only by a small total amount over that range.\cite{internal-resonator}  So, effectively, that opening is ``small."  For here, it is imagined that the cavities are isolated from each other. 

The ideal head modes are the standard modes of the full-radius drum head.  Again, linearity, superposition, and impedance mis-match make the physics tractable.  The head modes can be decomposed into modes suitable for driving the top surfaces of the two pot air volumes.  Each component generates a back reaction, which acts as an additional driver on the head --- all easy to imagine in all aspects qualitatively but a lot to implement quantitatively.  

Two further approximations help.  The annulus is unrolled into a rectangular box, as in FIG.~9.  The end boundary conditions are periodic.\cite{periodic}  A calculation for a rectangular box gives a good match to the observed resonances of an actual annulus.\cite{internal-resonator}  There are always an even number of  azimuthal node lines of the head as it drives the annulus.  These are preserved as the $x$-$z$ node planes, an even number of them evenly spaced in $y$.   The radial motion is fixed at $z=0$, which corresponds to the outer edge of the head.  Rather than considering the Bessel functions appropriate to the radial dependence, I simply restrict attention to frequencies sufficiently low that the radial dependence of the zeroth order driving by the head over the width $W$ of the annulus is simply linear.  

The driving top surface of the rectangular box provides the boundary condition

\centerline{$v_x(H,y,z,t) = A \; z \;  \text{cos}\,{{2\pi} \over {L}}n  y  \; \text{sin}\,\omega t$}
\noindent where $2n$ is the number of azimuthal nodes (i.e., $n=0$,1,2,3...).  

The function $z$ is extended beyond the physical region $0<z<W$ to $f(z) = z$ for $2mW<z<(2m+1)W$ for integer $m$.  In  the intervening intervals $f(z) = 2W - z$.  $f(z)$ has period $2W$ and is expanded in cosines:

\centerline{$f(z) = \sum_{j=0}^\infty  a_j \; \text{cos}\,({{ \pi} \over { 2W}}jz)$}

\noindent with

\centerline{$a_0 = \pi/ 4$}

\noindent and for $j \ne 0$

\centerline{$a_j = $ {\Large ${8\over {\pi j^2}}$} $\text{cos}\,${\Large ${{j\pi}\over2}$} $\text{sin}${\Large $({j\pi \over 4})^2$} .}
\medskip
\noindent In this form, $a_j$ has many zeros: $a_j=0$ for all odd $j$ and for $j=$ 4, 8, 12, 16, ...  For convenience, one can write the $j$'s for the non-zero $a_j$'s as $j=0$ and $j=4l+2$ with $l=0,1,2,3...$ to facilitate a sum over $j$.

%{\small \bf SEE MATHEMATICA \& ASK WHY IS THE PRE-FACTOR WHAT IT IS???  IT LOOKS LIKE IT GIVES THE RIGHT FUNCTION.}

The equations for {\bf v}$(x,y,z,t)$ and $p(x,y,z,t)$ look just as they did for 1-2D when expressed in vector form.  The desired solution that fits the driven boundary with the above $v_x(H,y,z,t)$ is a sum over $j$ with coefficients $a_j$ of solutions of the form

\centerline{${v}_x = B\,k_x\, \text{sin}\,k_x x \; \text{cos}\,k_y y \; \text{cos}\,k_z z \; \text{sin}\,\omega t$}

\centerline{${v}_y = B\,k_y\, \text{cos}\,k_x x \; \text{sin}\,k_y y \; \text{cos}\,k_z z \; \text{sin}\,\omega t$}

\centerline{${v}_z = B\,k_x\, \text{cos}\,k_x x \; \text{cos}\,k_y y \; \text{sin}\,k_z z \; \text{sin}\,\omega t$}

\medskip

\centerline{$p = \rho B \omega \, \text{cos}\,k_x x \; \text{cos}\,k_y y \; \text{cos}\,k_z z \; \text{cos}\,\omega t$ .}

\medskip

\noindent $k_y = {{2\pi} \over L}n$ with $n$ fixed by the driving force;  $k_z = {\pi \over {2W}}j$;  $k_x = \sqrt{{\omega^2 \over c^2} - k_y^2 - k_z^2}$;  and, when $k_x < 0$, the trig functions become hyperbolic. 

In the sum over $j$, the 3D wave vector {\bf k} takes on a series of values.  The squares of the components always sum to $\omega^2/c^2$.   Periodic boundary conditions in $y$ reflect the original rotational symmetry of the round pot.    That symmetry manifests as a translational symmetry in $y$.  An important consequence is that any given, particular sinusoidal structure of the head is preserved in the back reaction.  (For quantum waves, this is recognized as conservation of angular momentum.)  So $k_y$ is fixed by the boundary condition on {\bf v}$_x(H,y,z,t)$; it characterizes a particular number, $n$, of half waves in the $y$-direction.  $k_z$ appears in the sum as $j$ multiples of half-waves that fit in $z$.  That comes from the Fourier expansion of the function $z$ that is the representation of the radial direction on the original circular head.  Furthermore, because there is a boundary in the radial or $z$ direction, there will, in general, be mixing of different wave numbers in that direction under the action of the dynamics.   $k_x$ is whatever it has to be (including imaginary) to satisfy $\omega$.

How the head responds to the calculated air pressure (even restricted to the region bounding the internal annulus) depends on the whole head's dynamics and how it is being forced over its entirety.  So, I display here simply a particular aspect of the back reaction pressure:

\centerline{$p(H,y,W,t) = \rho\,A\,\bar P\;\text{cos}\,k_y y\;\text{cos}\,\omega t$}

\noindent which defines a function $\bar P$.  It is the sum of terms with resonance zeroes in the denominator:

\centerline{$\bar P = \omega \, \sum^\infty_{j=0}${\Large ${a_j \over {k_{x,j}\;\text{tan}\,k_{x,j}H}}$} ,}

\noindent and it is relatively featureless except for the location in $\omega$ of those zeroes.  It does not look anything like anything one might hear or measure on a real instrument because it makes no reference to string or head dynamics.  It simply gives an accounting of the pressure at the top of the internal resonator due to motion of the head in that region.  The input for that motion is of a constant amplitude as a function of frequency. Also, motion on other regions of the head will contribute to pressure at that location to the same order of the calculation and would, in practice, have to be added in to find the back reaction motion.

%For the particular values $L=2\pi$, $H=\pi/10, $W=\pi/20, and $c=1$, FIG.~13 is a plot of $\bar P$ versus $\omega$.  A small imaginary part is added to $\omega$.  Now complex, $\bar P$ is added to a much larger real constant, the absolute value is taken, and the larger real constant is subtracted.  That procedure captures the magnitude of $\bar P$ while retailing information about its relative sign. 

%\bigskip

\subsection{Helmholtz resonance}

The simple formula always given for the frequency of a Helmholtz resonator is

\medskip

\centerline{$\omega_H = c$ {\large  $\sqrt{{A_{\text{neck}}^2 \over {V \; V_{\text{neck}}}}}$} .}

\medskip

\noindent $\rho \, V_{\text{neck}}$ is the mass of the little plug of air oscillating in the neck.  So we can identify  Hooke's constant for the ``spring of the air"\cite{hooke} as $k_H =  \rho \, c^2 \, A_{\text{neck}}^2/V$, where $V$ is the volume of the resonator and $A_{\text{neck}}$ is the area of the interface of the neck with the main volume $V$.  In this approximation, any further details of the geometry are irrelevant.  The air inside $V$ expands and contracts uniformly at $\omega$, pushing the neck air volume in and out.  The actual flow pattern is necessarily complicated, particularly near the two ends of the neck.  Elaborate and careful numerical integration of the air equations of motion have identified this behavior.  But the simple picture is obviously a limit: all dimensions of $V$ must be much larger than those of the neck.  Otherwise there is necessarily some shape dependence.  Furthermore, the effective values of the parameters in the expression for $\omega_H$ cannot be deduced with precision simply by measuring the geometry of the resonator.  The hardest part is the ``neck."  On a real instrument, it is the outlet of the cavity volume to the ambient air.  Rather, effective values can be estimated and then refined by combinations of acoustic and mechanical manipulations.\cite{politzer}  Nevertheless, the simple picture is totally appropriate for the present endeavor.  It maps very simply onto the 0-0D model.  The cavity Helmholtz resonance provides a small back reaction onto certain head motions. 

\begin{figure}[h!]
\includegraphics[width=4.7in]{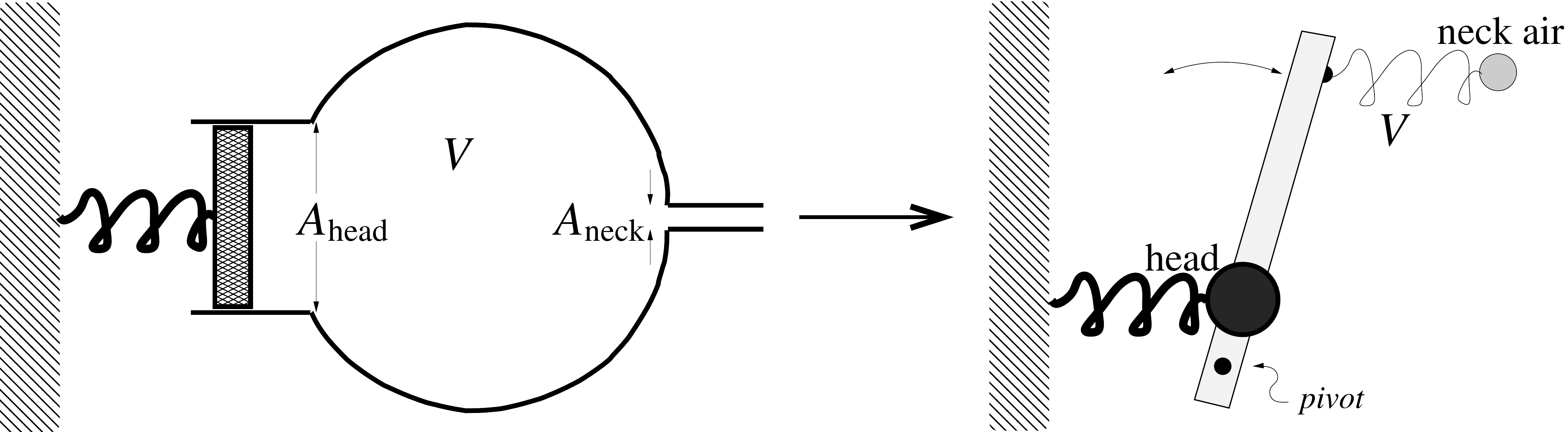}
 \caption{A driven Helmholtz resonator mapped onto a 0-0D two-oscillator system, emphasizing the necessary translation of the head motion measure into the measure of neck air motion} 
 \end{figure}

Any head motion that changes the volume of the cavity will excite the Helmholtz resonance.  The resonant amplitude depends on frequency in the standard way for a forced oscillator with resonant frequency $\omega_H$.  The back reaction on the head is a pressure that oscillates in time at the driving frequency but is constant over the area of the head.  Thus, irrespective of the particular driving shape, all head modes that change volume incur a back reaction motion.

For a circular drum head, only the axially symmetrical modes, i.e., with no diameter node lines, change the total volume.  Therefore only the purely radial modes are relevant.  These are indexed by the zeroes, $\alpha_{0,n}$, of the $0^{th}$ order Bessel function, $J_0$.  The $n^{th}$ zero matches the fixed head at the outer radius $R$, and there are $n-1$ other nodal circles interior to that.  So the radial profile is $J_0({{\alpha_{0,n}} \over R} r)$.  The total volume change decreases with $n$.  With a rectangular head, only modes with an even number of nodes in each direction give a net volume change.

Driving any one of the purely radial mode shapes will excite the Helmholtz mode.  The only change from the 0-0D model analyzed earlier is a further required calculation of a hydraulic effect, as suggest by FIG.~13.  The maximum zeroth order driving head motion displaces a certain amount of air.   That translates into a particular maximum amplitude for the motion of the neck air plug were there no compression and expansion of the air.  That geometric relation is the map of the problem onto the 0-0D springs.  The neck plug moves in response to the head moving the other end of its ``air spring."    The amplitude of the resulting plug motion corresponds to a pressure variation in $V$.  That oscillating pressure is uniform over the head.  Put back into the head equation of motion, this effects all of the purely radial modes to greater or lesser extent.

\subsection{2-3D}

At the level of this investigation, there is not much to add to the qualitative comments of section II.6.  For a rectangular banjo, the head mode shapes are sines in both directions.  Both must be expanded in cosines, which then gives a double sum as solutions for the air motion.  Its qualitative behavior is like the 1-2D case, except that there are more resonances per interval of driving frequency.  A  circular head and pot have rotational symmetry,  which, as discussed for the internal resonator, preserve the azimuthal structure of head modes in the air motion and back reaction.  However, the radial boundary requires an expansion in Bessel functions to get a solution for any particular driving mode.

\section{conclusions}

The goal of this investigation was to find the most elementary way to follow the physics of how of the air modes of the pot effect the motion of the head.  For the banjo, one wants a way to contemplate the whole possible range of frequency, pot design, and head excitation.  I do not know of anything simpler than what is presented.  To place this knowledge in the bigger picture, it is worth remembering that a frequency-domain, steady-state analysis, as employed here, accounts for only part of what's going on in the production of music.  In real time, typically only a small portion of the energy delivered to some system goes into sound.  Most goes into heat or is transferred to some other system.  On a banjo, the head is by far the most efficient transducer of vibration to sound.  So giving some of its energy to another subsystem is a losing proposition in terms of total sound production.  Nevertheless, we like the effects of the air modes and other chosen design elements because of the ways they spread out the responsiveness of the head to driving by the strings --- different designs to different degrees in different directions, according to taste.  If the head's response to string driving at a particular frequency is sub-optimal (whatever that is), the string's energy stays longer in the string.  The string makes virtually no sound on its own but does dissipate energy to friction.  The poor head response might prolong the sustain (albeit quieter) of the particular component of a pluck, but, more significantly, it results in more total energy loss in the string itself.  So the mixing up of frequencies and wavelengths contributes to shaping the sound according to our pleasure.  Of course, what we like is strongly influenced by what we know and what we're used to.  Musical instrument design is not, by and large, a purely engineering endeavor.


\begin{thebibliography}{99}

\bibitem{rayleigh}Rayleigh (J.~W.~Strutt), {\it The Theory of Sound}, $2^{\text{nd}}$ ed., vol. 1, \S213, Macmillan, London (1894)

\bibitem{rossing}e.g., {\it Effects of air loading on timpani membrane vibrations}, R.~Christian et al., J.Acoust.Soc.Am.,  v.76, no.5, November 1984, p.1337

\bibitem{kergomard}{\it Acoustics of Musical Instruments,} A.~Chaigne \& J.~Kergomard, Springer (2016, English edition), \S14.2, p.773-796

\bibitem{politzer}\href{http://www.its.caltech.edu/~politzer}{http://www.its.caltech.edu/\url{~}politzer}; D.~Politzer, {\it Physics of the Bacon Internal Resonator Banjo}, HDP:16-02 June 2016; {\it Banjo Rim Height and Sound in the Pot}, HDP:16-03 July 2016; and {\it The Open Back of the Open-Back Banjo}, HDP: 13-02 December 2013.

\bibitem{deutsch}e.g., S.~McAdams, {\it Muiscal Timbre Preception}, Ch.2 in {\it The Psychology of Music, Third Edition}, ed. D. Deutsch, Academic Press (2012)

\bibitem{internal-resonator}See ref. \cite{politzer} for description, analysis, sound samples of the internal resonator

\bibitem{mike-gregory}Rayleigh\cite{rayleigh} discusses rectangular drums and summarizes previous work.  The doshpuluur of Tuva is a trapezoidal fretless banjos.  With frets, one can even play bluegrass on a rectangular banjo, e.g., the renowned Johnny Butten on a ``Squared Eel" banjo: \href{https://www.youtube.com/watch?v=rYUoIewJpro}{https://www.youtube.com/watch?v=rYUoIewJpro}   .  That instrument is one of a whole line produced by Mike Gregory: \href{https://www.youtube.com/watch?v=_YTRzKr538Q&t=189s}{https://www.youtube.com/watch?v=\_YTRzKr538Q\&t=189s} .

\bibitem{greens}Of course, there is usually an equivalence between various methods. The construction presented here can be understood as the representation of the relevant Green's function as the sum of the outer products of modes, which yields the overlap of the mode with the problem at hand times that mode, then summed over the complete set of modes.

\bibitem{periodic}One may ask, ``Where is $y=0$ if the pot is cylindrical with perfect rotational symmetry?"  The at-first-confusing answer is, ``anywhere."  More precisely, if a zero is chosen arbitrarily, there will be periodic sine  and periodic cosine solutions with the same wavelength and frequency.  Any other zero location can be described by some appropriate superposition of the original sines and cosines.  From an equivalent perspective, there can be traveling waves of the same frequency and wavelength going one way around and going the other.  Any slight imperfection distinguishes the physically appropriate distinct solutions of (different) definite frequency.  This feature of the pot air is shared by the head modes and, in fact, any waves with an $O(2)$ symmetry.  The plucked string, tubular wind chimes, and Tibetan singing bowls are all examples with doubly degenerate normal modes that are split in frequency by small imperfections.  The three dimensional analog (with $SU(2)$) has been studied in quantum mechanics to great effect, e.g., atomic spectra in magnetic fields.  Interactions as well as small imperfections can make the proper denumeration of the states relevant.  Even if there is perfect symmetry, correct counting is sometimes essential, e.g., in thermal physics.

\bibitem{hooke} R. Hooke, as described by R. Boyle in {\it New Experiments Physico-Mechanicall, Touching the Spring of the Air, and its Effects...}, 1660.

\end{thebibliography}
\end{document}